\pdfoutput=1
\documentclass[11pt,letterpaper]{article}
\usepackage[letterpaper,margin=1in]{geometry}
\usepackage{amsmath,amssymb,amsthm}
\usepackage[T1]{fontenc}
\usepackage{lmodern}
\usepackage{xcolor}
\usepackage{graphicx}
\usepackage{tikz}
\usetikzlibrary{arrows.meta,positioning,fit,backgrounds}
\usepackage[colorlinks,citecolor=blue!60!black,linkcolor=blue!50!black,urlcolor=blue!50!black]{hyperref}

\newcommand{\class}[1]{\ensuremath{\mathbf{#1}}}
\renewcommand{\P}{\class{P}}
\newcommand{\FP}{\class{FP}}
\newcommand{\NP}{\class{NP}}
\newcommand{\ONP}{\class{ONP}}

\newcommand{\SOUP}{\class{SOUP}}
\newcommand{\PH}{\class{PH}}
\newcommand{\PP}{\class{PP}}
\newcommand{\BPP}{\class{BPP}}
\newcommand{\BPPpath}{\class{BPP_{path}}}
\newcommand{\ZPP}{\class{ZPP}}
\newcommand{\MA}{\class{MA}}
\newcommand{\AM}{\class{AM}}
\newcommand{\OMA}{\class{OMA}}
\newcommand{\SBP}{\class{SBP}}
\newcommand{\YP}{\class{YP}}
\newcommand{\E}{\class{E}}

\newcommand{\SIZE}{\class{SIZE}}
\newcommand{\TALLY}{\class{TALLY}}
\newcommand{\OMIN}{\class{OMIN}}
\newcommand{\OGTD}{\class{OGTD}}
\newcommand{\OUGTD}{\class{OUGTD}}
\newcommand{\OM}{\class{OM}}
\newcommand{\OUM}{\class{OUM}}
\newcommand{\Ppoly}{\class{P/poly}}
\newcommand{\co}{\class{co}\textbf{-}}
\newcommand{\LtP}{\class{L_2^{\class{\scriptscriptstyle P}}}}
\newcommand{\OLtP}{\class{OL_2^{\class{\scriptscriptstyle P}}}}
\newcommand{\StP}{\class{S_2^{\class{\scriptscriptstyle P}}}}
\newcommand{\OtP}{\class{O_2^{\class{\scriptscriptstyle P}}}}
\newcommand{\StE}{\class{S_2^{\class{\scriptscriptstyle E}}}}
\newcommand{\LtE}{\class{L_2^{\class{\scriptscriptstyle E}}}}
\newcommand{\SigmakP}[1]{\class{\Sigma_{#1}^{\class{\scriptscriptstyle P}}}}
\newcommand{\PikP}[1]{\class{\Pi_{#1}^{\class{\scriptscriptstyle P}}}}
\newcommand{\prStP}{\class{prS_2^{\class{\scriptscriptstyle P}}}}
\newcommand{\prOtP}{\class{prO_2^{\class{\scriptscriptstyle P}}}}
\newcommand{\prSBP}{\class{prSBP}}
\newcommand{\prAM}{\class{prAM}}
\newcommand{\prMA}{\class{prMA}}
\newcommand{\prOMA}{\class{prOMA}}
\newcommand{\OprMA}{\class{OprMA}}
\newcommand{\sOprMA}{\class{sOprMA}}
\newcommand{\prONP}{\class{prONP}}
\newcommand{\problem}[1]{\ensuremath{\textup{\textsc{#1}}}}
\newcommand{\SSE}{\problem{SSE}}
\newcommand{\SAT}{\problem{SAT}}
\newcommand{\LOP}{\problem{LOP}}

\newcommand{\bit}{\{0,1\}}
\newcommand{\YES}{\mathrm{YES}}
\newcommand{\NO}{\mathrm{NO}}
\newcommand{\out}{\mathrm{out}}
\newcommand{\Out}{\mathrm{Out}}
\DeclareMathOperator{\dg}{deg}
\newcommand{\eps}{\varepsilon}
\newcommand{\etalchar}[1]{$^{#1}$}
\newcounter{knownfact}

\newcommand{\factitem}[1]{\item[(F\number\numexpr\value{knownfact}+1\relax)]\refstepcounter{knownfact}\label{#1}}

\newtheorem{theorem}{Theorem}[section]
\newtheorem{lemma}[theorem]{Lemma}
\newtheorem{corollary}[theorem]{Corollary}
\newtheorem{proposition}[theorem]{Proposition}
\newtheorem{observation}[theorem]{Observation}

\theoremstyle{definition}
\newtheorem{definition}[theorem]{Definition}

\newtheorem*{claim}{Claim}
\theoremstyle{remark}
\newtheorem{remark}[theorem]{Remark}
\theoremstyle{plain}
\newtheorem{mainthm}{Theorem}

\title{The Ultimate Karp--Lipton Collapse}

\author{Edward A. Hirsch\thanks{Department of Computer Science, Ariel University, Israel.
Email: \texttt{edwardh@ariel.ac.il}. This research was conducted with the support of the
State of Israel, the Ministry of Immigrant Absorption, and the Center for the Absorption
of Scientists.%
}
\and
Ilya Volkovich\thanks{Computer Science Department, Boston College, Chestnut Hill, MA.
Email: \texttt{ilya.volkovich@bc.edu}.}}

\date{November 4, 2026}

\begin{document}
\maketitle

\begin{abstract}
Karp and Lipton~\cite{KL80} proved that if $\NP\subseteq\Ppoly$, then the polynomial
hierarchy collapses. A long line of work pushed the collapse to smaller classes.
In particular, Chakaravarthy and Roy obtained two such collapses, to (apparently) incomparable
classes $\OtP$~\cite{CR06} and 
$\P^{\prMA}$~\cite{CR11}.
Hirsch and Volkovich~\cite{HV26} united them by a collapse to
the input-oblivious version $\P^{\prOMA}$ of $\P^{\prMA}$, which is contained in both;
Hirsch, Volkovich and Williams~\cite{HVW26} obtained a collapse to a slightly smaller
class. Currently, the three
deepest known collapses, pairwise incomparable to the best of our current knowledge, 
are to $\P^{\prOMA}$ and its versions; to $\ZPP^{\ONP}$, the input-oblivious version of the collapse to $\ZPP^{\NP}$ of
K\"obler and Watanabe~\cite{KW98}, which builds on~\cite{BCG+96}; and to
$\OMIN=\OLtP$~\cite{HV26,HVW26}, the input-oblivious version of the class $\LtP$ of
Korten and Pitassi~\cite{KP24}. The last collapse is proved in this paper. The class $\OMIN$
has fixed-polynomial circuit lower bounds~\cite{HVW26}; no smaller collapse class with
such lower bounds is known.

We unite the three collapses in the same spirit, by a collapse to a class contained in
all of them. 
Derandomization results for input-oblivious classes in~\cite{HV26,HVW26}
in fact use a two-stage model: first the computation uses one oracle
(and ignores the input), then another oracle.
This model is inspired by the
untrusted advice classes of Aaronson and Drucker (including $\YP^*$)~\cite{AD14} and by the learning
algorithm of~\cite{BCG+96} as used in~\cite{KW98}. 
We use three-stage \emph{generator--tester--decoder} computations.
In this model, a polynomial-time
generator with an oracle computes an advice string from $1^n$, a tester certifies it,
and a polynomial-time decoder uses it on the input. 
Let $\OM$ be the class
of languages decided in this model by a generator with an oracle for a restricted
version of $\prOMA$, a $\co\ONP$ tester, and a $\P$ decoder. 
This class is contained in the three collapse classes above, $\OM\subseteq\OMIN\cap\P^{\prOMA}\cap\ZPP^{\ONP}$. We argue that it is a natural class (for example, closed under polynomial-time Turing reductions).
Each of the three
containments comes from executing the generator in a different way: with its oracle;
with its oracle replaced by a fixed language in $\LtP$; or with the oracle answers
sampled at random, after which the tester checks the advice. 
We show that if
$\NP\subseteq\Ppoly$, then $\PH=\OM$.
The collapse also holds with a unique certified advice string.

At the other end of the picture, all these input-oblivious collapse classes lie in
$\P^{\prSBP}$, and so
do even the non-input-oblivious versions $\P^{\prMA}$, $\LtP$ and $\StP$ of
$\P^{\prOMA}$, $\OLtP$ and $\OtP$, as well as $\P^{\prStP}$: we prove
$\P^{\prStP}\subseteq\P^{\prSBP}$. This
strengthens $\StP\subseteq\P^{\prAM}$ of Chakaravarthy and Roy~\cite{CR11} and our
earlier containment $\LtP\subseteq\P^{\prSBP}$~\cite{HV26}. The proof unites the
Goldreich--Wigderson-type  \cite{GW99} procedure of~\cite{CR11} with the approximate-counting
technique of~\cite{HV26}. Its input-oblivious counterpart concerns $\OtP$, the
first collapse class of Chakaravarthy and Roy: every language in $\OtP$ has
polynomial-size circuits constructible in $\FP^{\prSBP}$.
\end{abstract}

\thispagestyle{empty}
\newpage
\tableofcontents
\newpage
\section{Introduction}\label{sec:intro}

\subsection{Karp--Lipton collapses}

Karp and Lipton~\cite{KL80} proved that if $\NP\subseteq\Ppoly$, then the
polynomial hierarchy collapses to its second level (the second level is credited in~\cite{KL80} to
Sipser). Together with Kannan's
theorem~\cite{Kan82} that $\PH$ contains, for every $k$, a language without circuits
of size $n^k$, such a collapse yields \emph{fixed-polynomial circuit lower bounds} by a
\emph{win--win argument}. Namely, if $\NP\not\subseteq\Ppoly$, the lower bound holds already
for $\NP$; otherwise $\PH$ collapses to the class in question, which then inherits
Kannan's lower bound. A long line of work pushed the collapse to smaller classes:
$\ZPP^{\NP}$~\cite{BCG+96,KW98}; the symmetric alternation class
$\StP$~\cite{Can96,RS98}, by an argument of Sengupta (see~\cite{Cai07}); its
input-oblivious version $\OtP$~\cite{CR06}; and $\P^{\prMA}$~\cite{CR11}. Here $\prMA$
is the class of promise problems with Merlin--Arthur protocols; the letter
$\class{O}$ in the class name stands for \emph{input-obliviousness}: 
$\StP$ certificates (or Merlin's proofs in $\prMA$) may
depend on the length of the input but not on the input itself,
which defines the class $\OtP$ (or $\prOMA$, respectively).

While for a class containing $\NP$, a collapse immediately gives fixed-polynomial lower
bounds, input-oblivious classes lie in $\Ppoly$ by definition and therefore do not contain $\NP$ unless
$\NP\subseteq\Ppoly$; for them, the collapse and the lower bound are two separate
problems. For $\OtP$, a lower bound was found nearly two decades after the collapse,
by Gajulapalli, Li and Volkovich~\cite{GLV24}, using the range avoidance
problem~\cite{KKMP21,Kor22,CHR24,Li24}. In \cite{KP24}, Korten and Pitassi introduced the
class $\LtP$: an input is accepted when the minimum of a polynomial-time decidable
linear order, which depends on the input, starts with~$1$. In the same paper, they also proved
fixed-polynomial lower bounds for $\LtP$ and asked for a Karp--Lipton collapse.
Subsequently, Hirsch and Volkovich~\cite{HV26} showed the collapse by proving
$\P^{\prMA}\subseteq\LtP\subseteq\P^{\prSBP}$ and, in addition, introduced $\OLtP$ as the input-oblivious version of $\LtP$. Hirsch, Volkovich and Williams~\cite{HVW26} provided a different, yet equivalent, characterization of $\OLtP$ by showing that it 
equals the class $\OMIN$, in which the linear order itself depends only on the input length. They 
also obtained fixed-polynomial lower bounds for this class.

\paragraph{Uniting collapses.}
The two collapses of Chakaravarthy and Roy, to $\OtP$~\cite{CR06} and to\linebreak%
$\P^{\prMA}$~\cite{CR11}, are incomparable as far as we know: $\OtP$ is
input-oblivious, while $\P^{\prMA}$ contains $\NP$. Hirsch and Volkovich~\cite{HV26}
united them by deepening the collapse to $\P^{\prOMA}$, the input-oblivious version of
$\P^{\prMA}$. This class is contained in both: $\P^{\prOMA}\subseteq\P^{\prMA}$
trivially, and $\P^{\prOMA}\subseteq\OtP$~\cite{HV26}. They also observed that
input-oblivious collapses are currently the deepest ones. 
Hirsch, Volkovich and Williams proved that $\NP\subseteq\Ppoly$ implies
the collapse $\PH=\P^{\OprMA}$ to a slightly smaller
class~\cite[Proposition~6.9]{HVW26}, and their proof 
in fact gives a stricter promise class
$\sOprMA$ in which Merlin's oblivious proof dominates every other proof on
\emph{every} input and \emph{every} random string
(see Definition~\ref{def:soprma} for a formal definition).

\paragraph{Three deepest collapses.}
Three collapse classes are now the deepest known. As far as we know, they are
pairwise incomparable.
\begin{itemize}
\item $\P^{\prOMA}$ introduced in~\cite{HV26}, together with its slightly smaller versions
$\P^{\OprMA}$ \cite{HVW26} and $\P^{\sOprMA}$ (introduced in this paper).
\item $\ZPP^{\ONP}$, the input-oblivious version of the collapse to $\ZPP^{\NP}$ of
K\"obler and Watanabe~\cite{KW98}, which builds on~\cite{BCG+96}. It holds because
$\NP\subseteq\Ppoly$ implies $\NP=\ONP$~\cite{FSW09}.
\item $\OMIN=\OLtP$~\cite{HV26,HVW26}. We prove that $\NP\subseteq\Ppoly$ implies
$\PH=\OMIN$ (Theorem~\ref{thm:ominKL}). Since $\OMIN\not\subseteq\SIZE[n^k]$~\cite{HVW26},
no smaller collapse class with fixed-polynomial lower bounds is known. After the lower
bound of Gajulapalli, Li and Volkovich~\cite{GLV24}, the class $\OtP\supseteq\OLtP$ was for
a while the smallest natural input-oblivious class known to have both properties.
\end{itemize}
In the spirit of~\cite{HV26}, we unite these three collapses: we prove a collapse to a
class contained in all three of them (Theorem~\ref{thm:ultimate}).

A word of caution is due. A Karp--Lipton class inside all three classes above is easy
to obtain, even one of the generator--decoder kind described below. If $\mathcal C$ is
a Karp--Lipton class closed under polynomial-time Turing reductions, then so is its
\emph{tally (unary) closure} $\P^{\TALLY\cap\mathcal C}\subseteq\mathcal C$
(Theorem~\ref{thm:transfer}), in which the advice for length $n$ is computed from $1^n$
with an oracle from $\mathcal C$ (Proposition~\ref{prop:OP}(b)). For
$\mathcal C=\P^{\sOprMA}\cap\ZPP^{\ONP}$, the tally closure lies in $\P^{\sOprMA}$ and
$\ZPP^{\ONP}$ because it is contained in $\mathcal C$, and in
$\OMIN=\P^{\TALLY\cap\LtP}$ (Theorem~\ref{thm:OPL2}) because
$\mathcal C\subseteq\P^{\prMA}\subseteq\LtP$~\cite{HV26}. Such a class, however, is
obtained by a closure construction rather than from a natural model of computation. Our
class comes from such a model, described below.

\subsection{Generator--tester--decoder computations}

Our model has three components. They work as follows on inputs of length $n$.
\begin{itemize}
\item A \emph{generator} receives only $1^n$ (i.e., the input length instead of the input itself) and computes an advice string $a$.
It is a polynomial-time transducer with access to an oracle. If the oracle is a
promise problem, the access is \emph{``loose''}, and the generator may therefore produce several different
\emph{legal outputs}.
\item A \emph{tester} receives $(1^n,a)$. It must accept every legal output of the
generator, and it may accept only advice strings that work for the decoder.
\item A \emph{decoder} receives $(x,a)$ and decides whether $x\in L$.
\end{itemize}
We write $\OGTD[\mathcal A,\mathcal B,\mathcal C]$ (``oblivious
generator--tester--decoder'') for the class of languages decided this way, with a
generator from $\mathcal A$, a tester from $\mathcal B$, and a decoder from
$\mathcal C$ (Definition~\ref{def:OGTD}). We write $\ast$ when no tester is required,
$\OUGTD$ when the tester must accept exactly one advice string for each $n$, and
we abbreviate $\OGTD[\FP^{\mathcal X},\ast,\P]$ to $\OGTD[\mathcal X]$. Two-stage classes of this
kind have been studied in fact since Karp and Lipton's paper: their class $\P/\mathrm{poly}$
is $\OGTD[\mathrm{poly},\ast,\P]$, where the generator may be any polynomially bounded
advice function. K\"obler and Thierauf~\cite{KT94} bounded the complexity of
computing correct advice. For a language $A$, $\OGTD[A]$ is the class of languages
with $\P^A$-uniform polynomial-size circuits~\cite{Kan82,SW14}, and it equals
$\P^{\TALLY\cap\P^A}$ (Proposition~\ref{prop:OP}). See Section~\ref{sec:related} for more on
this line of work.

\medskip
The three-stage model is inspired by three predecessors.
\begin{itemize}
\item \emph{Untrusted advice.} The class $\YP^*$ of Aaronson and Drucker~\cite{AD14} is
$\OGTD[\mathrm{poly},\P,\P]$. There, the advice is tested in polynomial time without
access to the input.
\item \emph{Learning circuits.} In the algorithm of Bshouty, Cleve, Gavald\`a, Kannan, and
Tamon~\cite{BCG+96}, as used by K\"obler and Watanabe~\cite{KW98}, a $\ZPP^{\NP}$
learner generates circuits. Equivalence queries test them, and for $\SAT$ these tests
can be answered with an $\NP$ oracle.
\item \emph{Derandomization of oblivious classes.} The model underlies the classes
$\P^{\OLtP,\prONP}$ and $\P^{\OLtP,\ONP}$ used in~\cite{HV26,HVW26}, although it was
not stated explicitly there. In the proof of $\P^{\OprMA}\subseteq\P^{\OLtP,\ONP}$
\cite[Proposition~6.9]{HVW26}, the $\OLtP$ oracle generates the minimum $m$ of an
oblivious linear order that encodes a pseudorandom generator, and the rest of the
computation decodes. A Merlin--Arthur query $x$ becomes the $\ONP$ query ``some element
is smaller than $m$, or Merlin has a proof for $x$ that is accepted on the
pseudorandom strings encoded by $m$''. This query does not certify that $m$ is minimal:
it accepts every other candidate, and this disjunct makes the query input-oblivious;
the minimum, together with Merlin's oblivious proof, is a common witness. On the minimum, the query tests Merlin's
acceptance condition.
\end{itemize}

The main class of our paper defined along these lines is\footnote{The class name $\OM$ is
the common prefix of $\OMA$ and $\OMIN$: in $\OM$ the generator uses a (strict) oblivious
Merlin--Arthur oracle, and this class lies inside $\OMIN$.}
\[
 \OM=\OGTD[\FP^{\sOprMA},\co\ONP,\P],
\]
with the unambiguous-advice subclass $\OUM=\OUGTD[\FP^{\sOprMA},\co\ONP,\P]$. 

\subsection{Our results}\label{sec:results}

\begin{mainthm}[the ultimate Karp--Lipton collapse; Theorems~\ref{thm:three}
and~\ref{thm:OMcollapse}]\label{thm:ultimate}\ \\[-1.5em]
\begin{enumerate}
\item[\textup{(1)}] $\OUM\subseteq\OM\subseteq\OMIN\cap\P^{\sOprMA}\cap\ZPP^{\ONP}$.
\item[\textup{(2)}] If $\NP\subseteq\Ppoly$, then $\PH=\OM=\OUM$.
\end{enumerate}
\end{mainthm}

\begin{figure}[t]
\centering
\begin{tikzpicture}[x=2.2cm,y=1.55cm,
  every node/.style={font=\small},
  new/.style={-{Stealth[length=2.2mm]},very thick,blue!70!black},
  old/.style={-{Stealth[length=1.8mm]},gray!60!black},
  lab/.style={font=\scriptsize,midway,above,sloped,inner sep=1.5pt}]
\node (OUM) at (0,0) {$\OUM$};
\node (OM) at (1.5,0) {$\OM$};
\node[align=center] (OMIN) at (3.2,1.1) {$\OMIN=\OLtP$\\[-1pt]{\scriptsize fixed-polynomial lower bounds}};
\node (PS) at (3.2,0) {$\P^{\sOprMA}$};
\node (ZPP) at (3.2,-1.1) {$\ZPP^{\ONP}$};
\node (SOUP) at (3.2,2.25) {$\SOUP^{\ONP[1]}$};
\node (L2) at (5.4,1.1) {$\LtP$};
\node (POprMA) at (5.4,0) {$\P^{\OprMA}\subseteq\P^{\prOMA}$};
\node (ZPPNP) at (5.4,-1.1) {$\ZPP^{\NP}$};
\draw[old] (OUM) -- (OM);
\draw[new] (OM) -- (OMIN);
\draw[new] (OM) -- (PS);
\draw[new] (OM) -- (ZPP);
\draw[old] (OMIN) -- (L2);
\draw[old] (OMIN) -- node[font=\scriptsize,right,inner sep=2pt]{\cite{HVW26}} (SOUP);
\draw[old] (PS) -- (POprMA);
\draw[old] (ZPP) -- (ZPPNP);
\end{tikzpicture}
\caption{The ultimate collapse (Theorem~\ref{thm:ultimate}): If $\NP\subseteq\Ppoly$,
all these classes coincide with $\PH$. Thick arrows are the new unconditional
containments to the three previously known classes, two of them known
collapse classes; a collapse to $\OMIN$ is proved in this paper.}
\label{fig:ultimate}
\end{figure}
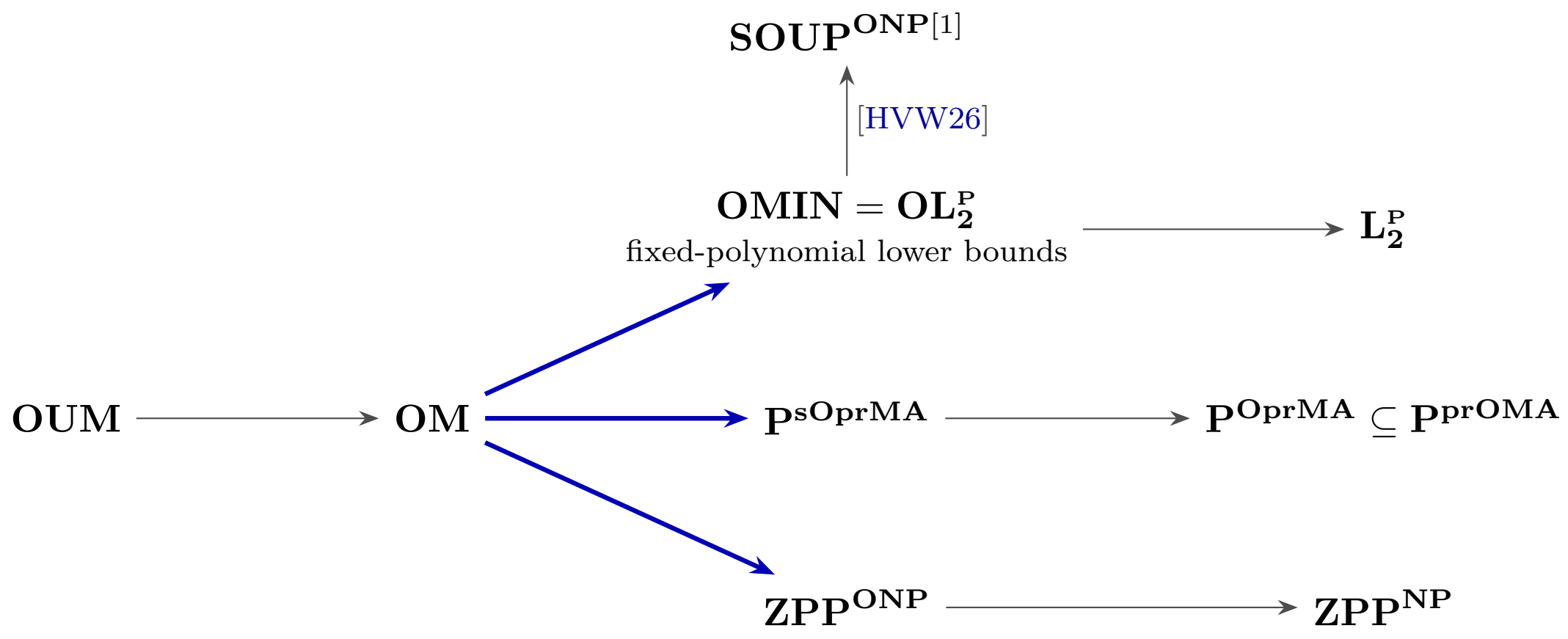

Each of the three containments in~(1) comes from executing the generator in a
different way.
\begin{itemize}
\item Running the generator with its oracle gives $\P^{\sOprMA}$.
\item Replacing the $\sOprMA$ oracle by a fixed language in $\LtP$ makes the advice
canonical, and the argument about tally oracles then gives $\OMIN$.
\item Sampling the generator with an $\ONP$ oracle and checking the result with the
tester gives $\ZPP^{\ONP}$. Sampling works because Merlin's proof dominates pointwise.
\end{itemize}
These containments are shown in Figure~\ref{fig:ultimate}, which also shows the upper bound $\OMIN\subseteq\SOUP^{\ONP[1]}$
of~\cite{HVW26}, where $\SOUP$ is their same-witness symmetric oblivious unambiguous class
(see Definition~\ref{def:soup}); thus
$\OM\subseteq\SOUP^{\ONP[1]}$ as well.
Fixed-polynomial lower bounds for $\OM$ remain open.

The collapse to $\OMIN$ is the main ingredient of the proof of
Theorem~\ref{thm:ultimate}, and it is of independent interest.

\begin{mainthm}[Corollary~\ref{cor:main-KL}]\label{thm:ominKL}
If $\NP\subseteq\Ppoly$, then $\PH=\OMIN=\OLtP$.
\end{mainthm}

As $\OLtP\subseteq\Ppoly$, Theorem~\ref{thm:ominKL} also implies that
$\NP\subseteq\Ppoly$ iff $\NP\subseteq\OLtP$ iff $\PH=\OLtP$
(Corollary~\ref{cor:equiv}). Unconditionally, it is open whether
$\ONP\subseteq\OLtP$.

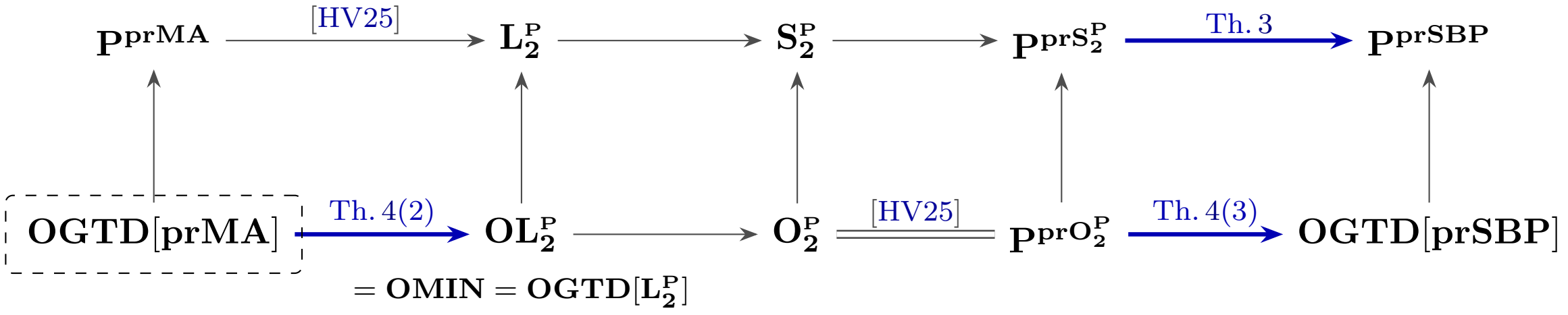
\begin{figure}[t]
\centering
\begin{tikzpicture}[x=2.25cm,y=1.9cm,
  every node/.style={font=\small},
  new/.style={-{Stealth[length=2.2mm]},very thick,blue!70!black},
  old/.style={-{Stealth[length=1.8mm]},gray!60!black},
  lab/.style={font=\scriptsize,midway,above,inner sep=1.5pt},
  nolb/.style={draw,dashed,rounded corners=2pt,inner sep=2pt}]
\node (PrMA)  at (0,1) {$\P^{\prMA}$};
\node (L2)    at (1.6,1) {$\LtP$};
\node (S2)    at (2.8,1) {$\StP$};
\node (PrS2)  at (3.95,1) {$\P^{\prStP}$};
\node (PrSBP) at (5.55,1) {$\P^{\prSBP}$};
\node (OPrMA)  at (0,0) {$\OGTD[\prMA]$};
\node (OL2)    at (1.6,0) {$\OLtP$};
\node (O2)     at (2.8,0) {$\OtP$};
\node (PrO2)   at (3.95,0) {$\P^{\prOtP}$};
\node (OPrSBP) at (5.55,0) {$\OGTD[\prSBP]$};
\node[font=\scriptsize,below=-1pt of OL2] {$=\OMIN=\OGTD[\LtP]$};
\draw[old] (PrMA) -- node[lab]{\cite{HV26}} (L2);
\draw[old] (L2) -- (S2);
\draw[old] (S2) -- (PrS2);
\draw[new] (PrS2) -- node[lab]{Th.\,\ref{thm:B}} (PrSBP);
\draw[new] (OPrMA) -- node[lab]{Th.\,\ref{thm:C}(2)} (OL2);
\draw[old] (OL2) -- (O2);
\draw[gray!60!black,line width=0.5pt,double,double distance=1.5pt] (O2) -- node[lab]{\cite{HV26}} (PrO2);
\draw[new] (PrO2) -- node[lab]{Th.\,\ref{thm:C}(3)} (OPrSBP);
\draw[old] (OPrMA) -- (PrMA);
\draw[old] (OL2) -- (L2);
\draw[old] (O2) -- (S2);
\draw[old] (PrO2) -- (PrS2);
\draw[old] (OPrSBP) -- (PrSBP);
\node[nolb,fit=(OPrMA)] {};
\end{tikzpicture}
\caption{All these classes coincide with $\PH$ unless $\NP\not\subseteq\Ppoly$.
Fixed-polynomial lower bounds are known for all of them \cite{CR11,KP24,GLV24,HVW26}
except for $\OGTD[\prMA]$ (dashed box).}
\label{fig:mirror}
\end{figure}

We also improve the upper bound for symmetric alternation, strengthening our earlier
upper bound for $\LtP$ \cite{HV26}, and then discuss what it means in the input-oblivious setting
(see Figure~\ref{fig:mirror}).

\begin{mainthm}[Theorem~\ref{thm:prS2}]\label{thm:B}
$\P^{\prStP}\subseteq\P^{\prSBP}$. In particular, $\StP\subseteq\P^{\prSBP}$.
\end{mainthm}

Here $\prSBP$ is the promise version of the class $\SBP$ of B\"ohler, Gla{\ss}er and
Meister~\cite{BGM06}, which lies between $\MA$ and $\AM$. A $\prSBP$ oracle is
equivalent to an oracle for approximate counting (see, e.g., \cite{HV26}). Moreover,
$\prSBP\subseteq\prAM$ and $\SBP\subseteq\BPPpath\subseteq\PP$~\cite{BGM06,HHT97}.
Theorem~\ref{thm:B} thus strengthens both $\StP\subseteq\P^{\prStP}\subseteq\P^{\prAM}$ of Chakaravarthy and
Roy~\cite{CR11} and the earlier containment $\LtP\subseteq\P^{\prSBP}$~\cite{HV26}.
In particular, $\P^{\prSBP}$ contains all the input-oblivious collapse classes discussed above. The three
deepest ones even lie in $\LtP\subseteq\P^{\prSBP}$~\cite{HV26}; for $\ZPP^{\ONP}$ this
follows from $\ZPP^{\ONP}\subseteq\BPP^{\ONP}\subseteq\P^{\OLtP,\ONP}\subseteq
\LtP$~\cite{HVW26}.
Together with
$\P^{\prMA}\subseteq\LtP\subseteq\P^{\prSBP}$~\cite{HV26}, Theorem~\ref{thm:B} shows
that $\P^{\prSBP}$ contains the non-input-oblivious versions $\P^{\prMA}$, $\LtP$ and
$\StP$ of $\P^{\prOMA}$, $\OLtP$ and $\OtP$, and even $\P^{\prStP}$. We do not know whether it
contains $\ZPP^{\NP}$ (Section~\ref{sec:open}). Theorem~\ref{thm:B} applies to every class inside
$\StP$, for instance to the classes $\class{PTW}$, $\class{PCW}$ and $\class{PMA}$ of
unambiguous $\SigmakP2$ problems defined by Gilboa, Goldberg, Koutsoupias and
Nisan~\cite{GGKN25}, who placed them in $\StP$. Since they also showed
$\LtP\subseteq\class{PTW}$, already the resulting containment
$\class{PTW}\subseteq\P^{\prSBP}$ strengthens $\LtP\subseteq\P^{\prSBP}$~\cite{HV26}. Its proof unites the techniques
of~\cite{CR11} and~\cite{HV26}. Chakaravarthy and Roy gave a version of the
Goldreich--Wigderson lemma \cite{GW99} for succinctly represented matrices that uses a $\prAM$ oracle.
We obtain one with approximate counting: given a circuit representing a Boolean matrix, a
polynomial-time algorithm with a $\prSBP$ oracle finds a small set of rows that
together beat every column (i.e., a \emph{row-side} collectively irrefutable certificate),
or a small set of columns that together beat every row (i.e., a \emph{column-side} collectively irrefutable certificate)
(Theorem~\ref{thm:gw}). As a special case, the same lemma gives another minimum-finding
algorithm, and thus another proof of $\LtP\subseteq\P^{\prSBP}$~\cite{HV26}
(Appendix~\ref{app:min}).

In the generator--decoder language, the
chain
$\P^{\prMA}\subseteq\LtP\subseteq\StP\subseteq\P^{\prSBP}$ has an input-oblivious
counterpart (the second line in Figure~\ref{fig:mirror}):

\begin{mainthm}[the oblivious setting; Theorems~\ref{thm:OPL2}, \ref{thm:OPprMA}
and~\ref{thm:O2}]\label{thm:C}\ \\[-1.5em]
\begin{enumerate}
\item[\textup{(1)}] $\OGTD[\LtP]=\OLtP\ (=\OMIN)$.
\item[\textup{(2)}] $\OGTD[\prMA]\subseteq\OLtP$.
\item[\textup{(3)}] $\OtP\subseteq\OGTD[\prSBP]$; that is, every language in $\OtP$
has polynomial-size circuits constructible in $\FP^{\prSBP}$.
\end{enumerate}
\end{mainthm}

Thus
\[
\begin{array}{ccccccc}
\P^{\prMA}&\subseteq&\LtP&\subseteq&\StP&\subseteq&\P^{\prSBP}\\[2pt]
\rotatebox{90}{$\subseteq$}&&\rotatebox{90}{$\subseteq$}&&\rotatebox{90}{$\subseteq$}&&\rotatebox{90}{$\subseteq$}\\[2pt]
\OGTD[\prMA]&\subseteq&\OLtP&\subseteq&\OtP&\subseteq&\OGTD[\prSBP]
\end{array}
\]
(see Figure~\ref{fig:mirror}), and Figure~\ref{fig:ultimate} places $\OM$ below the
three deepest collapse classes. Part~(3) concerns $\OtP$, one of the two collapse
classes united in~\cite{HV26}: its languages have polynomial-size circuits that can be
constructed with approximate counting.

\paragraph{Several byproducts.}
\begin{itemize}
\item \emph{Derandomizing approximate counting} (Corollary~\ref{cor:derand}). If
approximate counting can be done in $\FP^{\NP}$, then $\P^{\prStP}=\P^{\NP}$ and
$\OtP\subseteq\OGTD[\NP]$. Moreover, for every $k$, some language with polynomial-size
circuits constructible in $\FP^{\NP}$ has no circuits of size $n^k$. Previously,
to the best of our knowledge the collapse
$\StP=\P^{\NP}$ was known under a hardness assumption implied by the one used to
derandomize $\AM$~\cite{SU06}.
\item \emph{Direct proofs for order-based $\LtP$.} Korten and Pitassi~\cite{KP24} defined $\LtP$ via 
polynomial-time decidable linear orders
and stated that this definition is equivalent to the one by reductions to the linear ordering
principle~\cite[Theorems~5 and~6]{KP24}. To keep the paper self-contained, we use only the order-based
definition, which is also the one used in~\cite{HVW26}, and do not rely on this
equivalence anywhere
(Remark~\ref{rem:defs}). In particular, we prove closure under polynomial-time Turing
reductions for the order-based definition directly (Lemma~\ref{lem:closure}), and we
derive from it the pseudorandom generators that we use (Lemma~\ref{lem:prg}). Both definitions give subclasses of
$\StP$ (Remark~\ref{rem:defs}), and hence of $\P^{\prSBP}$ by Theorem~\ref{thm:B}.
\item \emph{Function classes and exponential time} (Appendix~\ref{sec:func}).
$\FP^{\prStP}\subseteq\FP^{\prSBP}$ and $\StE\subseteq\E^{\prSBP}$.
\end{itemize}

\subsection{Techniques}

\paragraph{Tally languages and canonical circuits (Theorem~\ref{thm:ominKL}).}
If $\NP\subseteq\Ppoly$, then every language $L\in\PH$ is polynomial-time reducible to
a tally (unary) language $T_L\in\PH$, which encodes the lexicographically first circuits for
$L$ (Lemma~\ref{lem:canonical}). Hence a Karp--Lipton collapse to $\P^{\mathcal C}$
transfers to $\OGTD[\mathcal C]$ (Theorem~\ref{thm:transfer}). Tally languages also
make linear orders oblivious: if $T\in\LtP$ is tally, its order
for $1^n$ depends only on $n$, so $T\in\OMIN$, and $\P^T\subseteq\OMIN$ since
$\P^{\OMIN}=\OMIN$~\cite{HVW26} (Lemma~\ref{lem:tally}). Finally, a $\prMA$ oracle can be replaced by an $\NP$
oracle together with a pseudorandom generator computed with an $\OLtP$
oracle~\cite{Kor22,KP24,HV26} (Lemma~\ref{lem:prg}). After this replacement the oracle is a language in
$\LtP$, so the advice becomes canonical and the tally argument applies
(Theorem~\ref{thm:OPprMA}). We work with the order-based definition of $\LtP$ and
prove its closure under polynomial-time Turing reductions directly
(Lemma~\ref{lem:closure}).

\paragraph{Certified minima (Theorem~\ref{thm:ultimate}).}
The minimum $\mu_n$ of an oblivious linear order is certified by the $\co\ONP$ test
``no element is smaller''. Indeed, $\mu_n$ itself is a common refutation of all other
candidates; this is also behind $\OMIN\subseteq\SOUP^{\ONP[1]}$~\cite{HVW26}. If
$\NP\subseteq\Ppoly$, every language in $\PH$ is in $\OMIN$ by
Theorem~\ref{thm:ominKL}. The bits of $\mu_n$ then form a tally language in
$\PH=\P^{\sOprMA}$. So an $\FP^{\sOprMA}$ generator computes $\mu_n$, a $\co\ONP$
tester certifies it, and the $\OMIN$ decoder decides. This puts $\PH$ into $\OUM$.
The three unconditional containments of $\OM$ come from simulating the generator:
directly, by derandomization inside $\LtP$, and by random sampling with an $\ONP$
oracle. The last one uses that, in $\sOprMA$, the set of pairs $(x,r)$ accepted with
\emph{some} proof is an $\ONP$ language.

\paragraph{Uniting the techniques of~\cite{CR11} and~\cite{HV26} (Theorem~\ref{thm:B}).}
Let $V(y,z)$ be a Boolean matrix given by a circuit. A row $y$ beats a column $z$ if
$V(y,z)=1$, and $z$ beats $y$ if $V(y,z)=0$. A set of rows is collectively irrefutable 
if every column is beaten by one of its rows, and similarly for columns. Goldreich and
Wigderson~\cite{GW99} observed that every matrix has a small collectively irrefutable
set of rows or of columns; if the matrix has an irrefutable row, the set must consist
of rows. Chakaravarthy and Roy~\cite{CR11} found such sets in $\FP^{\prAM}$ by a prefix
search guided by the \emph{best} extension of a prefix, an existential quantity. The
minimum-finding algorithm of~\cite{HV26} follows approximate \emph{averages} instead,
which a $\prSBP$ oracle provides. An above-average row, however, may beat few columns.
We therefore decide in each round whether to select a row or a column, by comparing
approximate counts of $0$- and $1$-entries of the remaining submatrix. On the denser
side, an above-average choice beats a quarter of the remaining lines.

\paragraph{Collective certificates (Theorem~\ref{thm:C}(3)).}
For $\ONP$ we use the \emph{domination game}. Its rows are candidate witnesses $w$, its
columns are pairs $(x,w')$, and $w$ loses only if $w'$ certifies $x$ while $w$ does
not. For $\OtP$ with verifier $V(x,y,z)$ and oblivious certificates $y_n,z_n$, we use
the \emph{consistency game}. Its rows are pairs $(y,z)$ and its columns are triples
$(x,z',y')$, and $(y,z)$ wins if $V(x,y,z')=1$ or $V(x,y',z)=0$. A collectively
irrefutable set of pairs yields yes-certificates that collectively beat every
no-certificate on every yes-instance. Relative to them, the complement of $L$ becomes an
$\ONP$-type language, and a second application of the lemma finds its witnesses.

\subsection{Related work}\label{sec:related}

\paragraph{Advice of bounded complexity.}
Karp and Lipton~\cite{KL80} defined advice classes $\mathcal C/\mathcal F$ for an
arbitrary class $\mathcal F$ of advice functions. Ko and Sch\"oning~\cite{KS85} bounded
the complexity of advice for sparse sets in the polynomial hierarchy. K\"amper's
nonuniform proof systems~\cite{Kam91} delimit the complexity of \emph{proof sets},
i.e., of sets of correct advice; in our terms, these are the sets accepted by a tester.
K\"obler and Thierauf~\cite{KT94} systematically restricted the complexity of
\emph{computing} correct advice. They also wrote $\mathcal C/\!/\mathcal F$ for advice
that depends on the input itself, which is why we avoid this notation. Gavald\`a~\cite{Gav95}
bounded the complexity of advice functions for sets in $\Ppoly$. Bshouty, Cleve,
Gavald\`a, Kannan and Tamon~\cite{BCG+96} showed that circuits can be learned exactly
with equivalence queries and an $\NP$ oracle in randomized expected polynomial time. As observed by K\"obler and
Watanabe~\cite{KW98}, correct advice for a language $A\in\Ppoly$ can then be computed
in $\class{FZPP}^{\NP^A}$ as a multi-valued function: different successful runs may
output different advice strings. Fortnow, Impagliazzo, Kabanets and
Umans~\cite{FIKU08} obtained this learning algorithm and $\StP\subseteq\ZPP^{\NP}$ from a
$\ZPP^{\NP}$ algorithm for succinct zero-sum games, and showed that approximating the
value of such a game within an additive error is complete for promise-$\StP$; by
Theorem~\ref{thm:B}, such approximations can be computed in $\FP^{\prSBP}$
(Appendix~\ref{sec:func}). Circuits constructible in a class $\mathcal C$, i.e.,
$\mathcal C$-uniform circuits, appear in Kannan's work~\cite{Kan82}; Santhanam and
Williams~\cite{SW14} proved that $\NP$ does not have $\P^{\NP}_{\|}$-uniform circuits of
size $n^k$. Marshall, Gyurik and Dunjko~\cite{MGD24} consider a related notion of
``bounded advice'', in which the advice must be computable within a given complexity
class.

\paragraph{Untrusted and verifiable advice.}
Aaronson~\cite{Aar07} introduced $\YP=\ONP\cap\co\ONP$, and Aaronson and
Drucker~\cite{AD14} introduced $\YP^*$, in which the advice is tested without the
input. Input-oblivious proof systems~\cite{FSW09,GoldreichMeir2015} can be viewed as
restricting $\Ppoly$ to verifiable advice. The same-witness class $\SOUP$ of~\cite{HVW26}
is an unambiguous version of $\YP^*$.

\paragraph{Other recent work.}
Chen, Hirahara and Ren~\cite{CHR24} proved that $\StE$ with one bit of advice requires
circuits of size $2^n/n$, and Li~\cite{Li24} removed the advice bit (see
also~\cite{CHLR26}). Ren and Williams~\cite{RW26} give
near-maximum circuit lower bounds for exponential-time classes with Merlin--Arthur
queries. Ghentiyala, Li and
Stephens-Davidowitz~\cite{GLS26} showed that decision problems reducible to range avoidance
lie in $\AM\cap\co\AM$. Gajulapalli, Ghentiyala, Li and Saraogi~\cite{GGLS25} placed the
linear ordering principle in $\class{UEOPL}^{\NP}$. Gilboa, Goldberg, Koutsoupias and
Nisan~\cite{GGKN25} defined syntactic classes of unambiguous $\SigmakP2$ problems and
placed them in $\StP$. Chen, McKay,
Murray and Williams~\cite{CMMW19} proved that fixed-polynomial lower bounds for
$\P^{\NP}$ are equivalent to a weak Karp--Lipton theorem. Shaltiel and
Umans~\cite{SU06} and Aydinlio\u{g}lu, Gutfreund, Hitchcock and Kawachi~\cite{AGHK11}
related the derandomization of approximate counting to hardness. On access to promise
oracles, see~\cite{ESY84,CHV92,GS88,BF99,CR11,HV26,MPR26}; the term \emph{``loose access''}
is from~\cite{HirschSokolov15} (while the first source using this concept was apparently \cite{CHV92}).

\subsection{Organization}\label{sec:organization}
Section~\ref{sec:prelim} gives definitions, including generator--tester--decoder
computations. Section~\ref{sec:orders} treats order-based $\LtP$ and proves the
collapse to $\OMIN$ (Theorem~\ref{thm:ominKL}) and parts (1) and~(2) of
Theorem~\ref{thm:C}. Section~\ref{sec:ultimate} proves the ultimate collapse
(Theorem~\ref{thm:ultimate}). Section~\ref{sec:gw} proves the
Goldreich--Wigderson lemma with approximate counting and Theorem~\ref{thm:B}, 
and Section~\ref{sec:mirror} applies it in the oblivious setting and proves
Theorem~\ref{thm:C}(3). Section~\ref{sec:open} lists open problems.
Appendix~\ref{app:routes} gives alternative routes to the collapses and discusses
relativization, and Appendix~\ref{sec:conseq} derives several byproducts
described at the end of Section~\ref{sec:results} above.

\section{Preliminaries}\label{sec:prelim}

\subsection{Conventions}
All polynomials are nondecreasing and at least~$1$. The letter $n$ denotes the length
of the input. In definitions based on polynomial-time predicates, every certificate,
proof, random string and advice string for inputs of length $n$ ranges over strings
of one fixed length, a polynomial in $n$ given by the definition ($p(n)$, or $a(n)$,
$b(n)$, \dots, when several lengths occur). Quantifiers such as $\exists y$, $\forall z$
and $\forall r$, and probabilities $\Pr_r$ (always over the uniform distribution), range
over strings of exactly this length. In particular, an input-oblivious certificate
$y_n$ has the same length as every other certificate for inputs of length~$n$.
Machines run in time polynomial in the length of their input, which is $n$ for a tally
input $1^n$; oracle machines are deterministic unless stated otherwise.

For a polynomial $p$ and
$n\in\mathbb N$, strings of length $p(n)$ are identified with tuples of strings of
prescribed lengths whenever these lengths depend only on~$n$. Pairs are encoded in a
length-regular way: the length of $(x,w)$ determines $|x|$ and $|w|$. A language $T$
is \emph{tally} if $T\subseteq\{1\}^*$; $\TALLY$ denotes the class of tally languages.
We use the Cantor pairing $\pi(n,i)=\binom{n+i+1}{2}+i$, a bijection
$\mathbb N^2\to\mathbb N$ such that $\pi$ and $\pi^{-1}$ are computable in time
polynomial in the unary lengths and $\pi(n,i)\le(n+i+1)^2$. We fix an encoding of
circuits with $n$ inputs and at most $s$ gates by strings of length exactly
$\ell(n,s)=O(s\log(n+s))$. Whether a string encodes such a circuit, and the value of
the circuit on a given input, are decidable in polynomial time. For a circuit $C$,
$\#C$ denotes the number of its satisfying assignments.

\subsection{Promise problems and oracle access}

A \emph{promise problem}~\cite{ESY84} is a pair $\Pi=(\Pi_\YES,\Pi_\NO)$ of disjoint
sets of strings; a language $L$ is identified with $(L,\overline L)$. A language $O$ is
\emph{consistent} with $\Pi$ if $\Pi_\YES\subseteq O$ and $\Pi_\NO\cap O=\emptyset$.

Oracle access to a promise problem $\Pi$ is \emph{loose}: queries in
$\Pi_\YES\cup\Pi_\NO$ are answered correctly, every other query receives an arbitrary
answer, and the computation must be correct for all possible answers. These answers
need not be consistent: the same query may receive different answers when it is asked
again, and, for a nondeterministic machine, on different computation paths. This mode
of access was introduced by Cai, Hemachandra and Vysko\v{c}~\cite[Definition~5.8]{CHV92}
as \emph{arbitrary failure access}, and it is the mode used by Chakaravarthy and
Roy~\cite{CR11}; the term \emph{loose} is from~\cite{HirschSokolov15}. For deterministic
polynomial-time oracle machines, loose access is equivalent to access through an
arbitrary language consistent with $\Pi$, as in Buhrman and Fortnow~\cite{BF99}. Indeed,
such a machine can store the answers it has received and never repeat a query, and then
every sequence of answers is given by some consistent language. We use the latter form
throughout, and we assume that deterministic oracle machines never repeat a query.

\begin{definition}
Let $\Pi$ be a promise problem. A language $L$ is in $\P^{\Pi}$ if there is a
deterministic polynomial-time oracle machine $M$ such that $M^{O}$ decides $L$ for
every language $O$ consistent with $\Pi$. A search problem (relation) $R$ is in
$\FP^{\Pi}$ if there is a deterministic polynomial-time oracle machine $M$ such that,
for every $O$ consistent with $\Pi$ and every input $x$, $M^O(x)$ is a solution
for~$x$. For a class $\mathcal C$ of promise problems,
$\P^{\mathcal C}=\bigcup_{\Pi\in\mathcal C}\P^{\Pi}$, and similarly for
$\FP^{\mathcal C}$.
\end{definition}

For $\Pi_1,\dots,\Pi_r$, let $\Pi_1\oplus\dots\oplus\Pi_r$ be the promise problem with
yes-instances $\langle i,q\rangle$ for $q\in(\Pi_i)_\YES$ and no-instances
$\langle i,q\rangle$ for $q\in(\Pi_i)_\NO$, where $i$ is written with a fixed number of
bits. A language consistent with it yields, by restriction, languages consistent with
each $\Pi_i$.

\subsection{Complexity classes}

\begin{definition}[$\prStP$, $\StP$~\cite{Can96,RS98}; $\prOtP$, $\OtP$~\cite{CR06}]
A promise problem $\Pi$ is in $\prStP$ if there are polynomials $a,b$ and a
polynomial-time predicate $V$ such that for every $x$,
\begin{align*}
 x\in\Pi_\YES&\Rightarrow\exists y\in\bit^{a(|x|)}\ \forall z\in\bit^{b(|x|)}\ \ V(x,y,z)=1,\\
 x\in\Pi_\NO&\Rightarrow\exists z\in\bit^{b(|x|)}\ \forall y\in\bit^{a(|x|)}\ \ V(x,y,z)=0.
\end{align*}
Such $y$ and $z$ are called \emph{irrefutable certificates}. $\Pi\in\prOtP$ if,
moreover, for every $n$ there are $y_n$ and $z_n$ that are irrefutable for all
$x\in\Pi_\YES\cap\bit^n$ and all $x\in\Pi_\NO\cap\bit^n$, respectively. $\StP$ and
$\OtP$ are the corresponding classes of languages.
\end{definition}

\begin{definition}[$\ONP$~\cite{FSW09,CR06}]
A language $L$ is in $\ONP$ if there are a polynomial $p$, a polynomial-time predicate
$V$, and strings $w_n\in\bit^{p(n)}$ such that, for every $x\in\bit^n$, $x\in L$
implies $V(x,w_n)=1$, and $x\notin L$ implies $V(x,w)=0$ for all $w$. We call $w_n$ an
\emph{oblivious witness}.
\end{definition}

\begin{definition}[$\prMA$, $\prOMA$~\cite{FSW09,HV26}; $\OprMA$~\cite{HVW26}; $\sOprMA$]
\label{def:soprma}
$\Pi\in\prMA$ if there are a polynomial $s$ and a polynomial-time predicate $V$ such
that for every $x$, with $y,r$ ranging over $\bit^{s(|x|)}$,
\begin{align*}
 x\in\Pi_\YES&\Rightarrow\exists y\ \forall r\ \ V(x,y,r)=1,\\
 x\in\Pi_\NO&\Rightarrow\forall y\ \ \Pr_r[V(x,y,r)=1]<1/4.
\end{align*}
For the input-oblivious versions, there are, in addition, strings $y_n\in\bit^{s(n)}$,
the \emph{oblivious proofs}, and we consider the following conditions on $x\in\bit^n$,
with $y,r$ ranging over $\bit^{s(n)}$:
\begin{align*}
 &x\in\Pi_\YES\ \Rightarrow\ \forall r\ \ V(x,y_n,r)=1,\tag{OMA1}\label{cond:OMA1}\\
 &x\in\Pi_\NO\ \Rightarrow\ \forall y\ \ \Pr_r[V(x,y,r)=1]<1/4,\tag{OMA2}\label{cond:OMA2}\\
 &x\notin\Pi_\NO\ \Rightarrow\ \forall y\ \forall r\ \ \bigl(V(x,y,r)=1\Rightarrow
  V(x,y_n,r)=1\bigr),\tag{OMA3}\label{cond:OMA3}\\
 &\forall y\ \forall r\ \ \bigl(V(x,y,r)=1\Rightarrow V(x,y_n,r)=1\bigr).\tag{OMA4}\label{cond:OMA4}
\end{align*}
$\Pi\in\prOMA$ if \eqref{cond:OMA1} and \eqref{cond:OMA2} hold for every $n$ and every $x\in\bit^n$.
$\Pi\in\OprMA$ if, in addition, \eqref{cond:OMA3} holds. $\Pi\in\sOprMA$ (strict $\OprMA$) if
\eqref{cond:OMA1}, \eqref{cond:OMA2} and \eqref{cond:OMA4} hold.
\end{definition}
Perfect completeness and soundness $1/4$ are without loss of generality for
$\prMA$~\cite{ZF87}. Since \eqref{cond:OMA4} implies \eqref{cond:OMA3},
$\sOprMA\subseteq\OprMA\subseteq\prOMA\subseteq\prMA$. Condition \eqref{cond:OMA3} says that the
oblivious proof $y_n$ dominates every other proof on every input outside $\Pi_\NO$ and
every random string; \eqref{cond:OMA4} requires this on every input. Every $\ONP$ language $L$ is a (total) problem in $\sOprMA$: take
$V(x,y,r)=V'(x,y)$ for an $\ONP$ verifier $V'$ of $L$.

\begin{definition}[$\SOUP$~\cite{HVW26}]\label{def:soup}
A language $L$ is in $\SOUP^A$ if there are a polynomial $p$, polynomial-time oracle
machines $Y,Z$, and strings $w_n\in\bit^{p(n)}$ such that, for every $x\in\bit^n$:
$Y^A(x,u)=Z^A(x,u)=0$ for every $u\neq w_n$; if $x\in L$, then $Y^A(x,w_n)=1$ and
$Z^A(x,u)=0$ for all $u$; and if $x\notin L$, then $Z^A(x,w_n)=1$ and $Y^A(x,u)=0$ for
all $u$. In $\SOUP^{\mathcal C[1]}$, the machines make one query to an oracle
from~$\mathcal C$.
\end{definition}

\paragraph{Approximate counting.}
The promise problem $\SSE$ (set size estimation) has yes-instances $(C,m)$ with
$\#C\ge m$ and no-instances $(C,m)$ with $\#C\le m/2$, where $C$ is a circuit and $m\ge1$
is an integer given in binary. It is complete for the class $\prSBP$ of B\"ohler, Gla{\ss}er and
Meister~\cite{BGM06} (see~\cite[Lemma~3.2]{HV26}), and we write
$\P^{\prSBP}=\P^{\SSE}$ and $\FP^{\prSBP}=\FP^{\SSE}$. Since
$\prSBP\subseteq\prAM$~\cite{BGM06}, $\P^{\prSBP}\subseteq\P^{\prAM}$. Note also that
$\SAT\in\P^{\SSE}$, via the query $(C,1)$.

\begin{lemma}[{\cite[Lemma~3.4]{HV26}}]\label{lem:count}
Given a circuit $C$ and a rational $\eps\in(0,1]$, one can compute an integer $t$ with
$\#C\le t\le(1+\eps)\#C$ in time polynomial in $|C|$ and $1/\eps$, with any oracle
consistent with $\SSE$. In particular, $t=0$ if and only if $\#C=0$.
\end{lemma}

\paragraph{Linear orders.}
Throughout this paper, $\LtP$ is the class defined by polynomial-time decidable linear
orders.

\begin{definition}[$\LtP$~\cite{KP24}]
\label{def:L2}
A language $L$ is in $\LtP$ if there are a polynomial $p$ and a polynomial-time
predicate $V$ such that, for every $x$, the relation $u\prec_x w\Leftrightarrow
V(x,u,w)=1$ is a strict linear order on $\bit^{p(|x|)}$ whose minimum starts with $1$
if $x\in L$ and with $0$ if $x\notin L$.
\end{definition}

\begin{remark}\label{rem:defs}
Korten and Pitassi~\cite{KP24} also consider the class of languages that are
polynomial-time (or $\P^{\NP}$-) reducible to the total search problem $\LOP$. Given a
circuit $E$ on pairs of strings, $\LOP$ asks for the minimum of the order defined by
$E$ or for a violation of the axioms of strict linear orders. They state that the two
definitions are equivalent~\cite[Theorems~5 and~6]{KP24}. To keep the paper
self-contained, we use only Definition~\ref{def:L2} and do not rely on this
equivalence anywhere. In particular, we prove closure under polynomial-time Turing
reductions for Definition~\ref{def:L2} directly (Lemma~\ref{lem:closure}) rather than
deriving it from the reduction-based definition, and we derive the pseudorandom
generators of \ref{fact:F3} from it (Lemma~\ref{lem:prg}). Both definitions give subclasses of
$\StP$, and hence of $\P^{\prSBP}$ by Theorem~\ref{thm:B}. For Definition~\ref{def:L2},
this is immediate. For reductions to $\LOP$, note that $\LOP$ has an $\class{F}\StP$
algorithm that determines a fixed solution for each instance~\cite[Lemma~5]{KP24}. The
bits of this solution form a language in $\StP$, and a reduction that works for every
solution of $\LOP$ works for this one. Hence these classes lie in
$\P^{\NP,\StP}=\P^{\StP}=\StP$ (see~\cite{CR11} for the last equality).
\end{remark}

\begin{definition}[$\OLtP$~\cite{HV26}; $\OMIN$~\cite{HVW26}]\label{def:OMIN}
$L\in\OLtP$ if $L\in\LtP$ via some $V,p$ such that for every $n$ there are strings
$y_n,z_n$ with $\min{\prec_x}=1y_n$ for all $x\in L\cap\bit^n$ and $\min{\prec_x}=0z_n$ for
all $x\in\bit^n\setminus L$. $L\in\OMIN$ if there are a polynomial $p$, a predicate
$U(1^n,u,w)$ computable in time polynomial in $n$ such that $u<_n
w\Leftrightarrow U(1^n,u,w)=1$ is a strict linear order on $\bit^{p(n)}$ for every
$n$, and a polynomial-time decoder $D$ such that $L(x)=D(x,\mu_{|x|})$ for every $x$,
where $\mu_n=\min{<_n}$.
\end{definition}

We use the following known facts.
\begin{itemize}
\factitem{fact:F1} $\NP\subseteq\LtP$~\cite{KP24}, with an explicit linear order for every
$\NP$ verifier~\cite[Lemma~2.11]{HVW26}.
\factitem{fact:F2} $\OLtP=\OMIN$~\cite[Theorem~3.4]{HVW26}, and $\OLtP\not\subseteq\SIZE[n^k]$
for every $k$~\cite[Theorem~3.3]{HVW26}. Clearly $\OLtP\subseteq\LtP$. Moreover,
$\OMIN\subseteq\OtP$. If $L\in\OMIN$ via $(p,<_n,D)$, the verifier
$V(x,y,z)=D(x,\min_{<_n}\{y,z\})$ has the irrefutable certificates $y_n=z_n=\mu_n$.
\factitem{fact:F3} (pseudorandom generators from oblivious orders) There are a language
$B\in\OLtP$ and a deterministic polynomial-time oracle algorithm that, given $1^N$,
outputs with oracle $B$ a list $g_1,\dots,g_m\in\bit^N$ with $m=\mathrm{poly}(N)$
such that every $N$-input circuit $C$ of size at most $N$ satisfies
$\bigl|\Pr_i[C(g_i)=1]-\Pr_{y\in\bit^N}[C(y)=1]\bigr|\le1/N$. This was observed
in~\cite[Proposition~5.5]{HV26} and~\cite[Lemma~6.2]{HVW26}, whose proofs pass through
the reduction-based definition of $\LtP$ (Remark~\ref{rem:defs}). Lemma~\ref{lem:prg}
derives \ref{fact:F3} from Definition~\ref{def:L2} alone.
\factitem{fact:F4} $\P^{\prOtP}=\OtP$~\cite[Theorem~4.3]{HV26},
$\P^{\prOMA}\subseteq\OtP$~\cite[Corollary~4.4]{HV26}, and, if $\NP\subseteq\Ppoly$,
then $\PH=\P^{\prOMA}$~\cite[Proposition~4.1]{HV26}. Moreover,
$\OtP\not\subseteq\SIZE[n^k]$ for every $k$~\cite{GLV24}.
\end{itemize}

All Karp--Lipton collapse classes considered in this paper, as well as $\P^{\prSBP}$ and
$\P^{\prAM}$, are contained in $\PH$. In
particular, $\OMIN\subseteq\SigmakP2\cap\PikP2$: for $x\in\bit^n$, $x\in L$ iff
$\exists m\,[\forall m'\ \neg(m'<_n m)\wedge D(x,m)=1]$, and similarly for $x\notin L$.
Moreover, $\P^{\prAM}\subseteq\P^{\SigmakP2}$. Indeed, a promise problem with an
Arthur--Merlin protocol with perfect completeness~\cite{FGM+89} is consistent with the
$\PikP2$ language of inputs $x$ such that, for every random string, Merlin has an
accepting reply.

\subsection{Generator--tester--decoder computations}

\begin{definition}[$\OGTD$]\label{def:OGTD}
Let $\mathcal A$ be a class of generators, $\mathcal B$ a class of languages
(testers), and $\mathcal C$ a class of languages (decoders). The generators we use are
deterministic polynomial-time oracle transducers $G$ with an oracle $\Pi$ from a fixed
class $\mathcal X$ of promise problems; this class of generators is denoted
$\FP^{\mathcal X}$. Thus $\mathcal A$ is a class of machines together with their oracles
(in our uses, $\FP^{\mathcal X}$ or the nonuniform class $\mathrm{poly}$ defined below),
whereas $\mathcal B$ and $\mathcal C$ are classes of languages (in our uses, $\co\ONP$,
$\P$ or $\ast$, and $\P$, respectively). On input $1^n$ the output of $G$ has a fixed length $p(n)$ for a
polynomial $p$. Its set of \emph{legal outputs} is
\[
 \Out_G(n)=\{G^O(1^n): O\text{ is a language consistent with }\Pi\}.
\]
A language $L$ belongs to $\OGTD[\mathcal A,\mathcal B,\mathcal C]$ if there are a
generator $G\in\mathcal A$, a tester $T\in\mathcal B$ (a language of pairs
$(1^n,a)$), and a decoder $D\in\mathcal C$ (a language of pairs $(x,a)$) with the
following properties. Informally, for inputs of length $n$, the generator computes an
advice string from $1^n$ alone; every advice string that the generator can output
passes the test, which does not see the input either; and every advice string that
passes the test lets the decoder decide $L$ correctly on all inputs of length $n$.
Formally, let
\[
 \mathrm{Acc}_T(n)=\{a:(1^n,a)\in T\}\quad\text{and}\quad
 \mathrm{Good}_D(n)=\bigl\{a\in\bit^{p(n)}:\ D(x,a)=L(x)\text{ for all }x\in\bit^n\bigr\}
\]
be the sets of \emph{accepted} and of \emph{good} advice strings, where we identify
languages with their characteristic functions. Then we require, for every $n$,
\begin{equation}\label{eq:OGTD}
 \Out_G(n)\ \subseteq\ \mathrm{Acc}_T(n)\ \subseteq\ \mathrm{Good}_D(n).
\end{equation}
$L$ belongs to $\OUGTD[\mathcal A,\mathcal B,\mathcal C]$ if, moreover,
$\mathrm{Acc}_T(n)$ has exactly one element for every $n$. We write $\ast$ for the
class of all languages. In $\OGTD[\mathcal A,\ast,\mathcal C]$ one can take
$\mathrm{Acc}_T(n)=\mathrm{Good}_D(n)$, so only $\Out_G(n)\subseteq\mathrm{Good}_D(n)$
is required. Finally, we abbreviate
\[
 \OGTD[\mathcal X]=\OGTD[\FP^{\mathcal X},\ast,\P];
\]
here the argument $\mathcal X$ is the class of oracles of the generator.
We also use the nonuniform generator class $\mathrm{poly}$ of all polynomially length
bounded functions $n\mapsto a_n$, whose only legal output on $1^n$ is $a_n$.
\end{definition}

For example, $\Ppoly=\OGTD[\mathrm{poly},\ast,\P]$, and $\YP^*=\OGTD[\mathrm{poly},\P,\P]$
by the definition of Aaronson and Drucker~\cite[Definition~18]{AD14}: there the tester
checks the advice without access to the input. A language $L$ is in $\OGTD[\Pi]$ if
and only if $L$ has polynomial-size circuits that can be constructed from $1^n$ in
$\FP^{\Pi}$. Indeed, $D(\cdot,a)$ can be written as a circuit, and conversely a circuit
can be evaluated. Since a generator is a deterministic polynomial-time machine, it
makes no difference whether its access to $\Pi$ is loose or through consistent
languages. Our main class is
\[
 \OM=\OGTD[\FP^{\sOprMA},\co\ONP,\P],\qquad
 \OUM=\OUGTD[\FP^{\sOprMA},\co\ONP,\P].
\]

\begin{proposition}\label{prop:OP}
Let $\Pi,\Pi',\Pi_1,\dots,\Pi_r$ be promise problems.
\begin{enumerate}
\item[\textup{(a)}] $\OGTD[\Pi]\subseteq\P^{\Pi}\cap\Ppoly$, and $\OGTD[\Pi]$ is closed
under complement.
\item[\textup{(b)}] For a language $A$, we have $\OGTD[A]=\P^{\TALLY\cap\P^{A}}$.
\item[\textup{(c)}] If $L\in\P^{L_1,\dots,L_r}$ and $L_i\in\OGTD[\Pi_i]$ for all $i$,
then $L\in\OGTD[\Pi_1\oplus\dots\oplus\Pi_r]$. In particular,
$\P^{\OGTD[\Pi]}=\OGTD[\Pi]$.
\item[\textup{(d)}] Suppose that there is a deterministic polynomial-time oracle
machine $M$ such that for every language $O$ consistent with $\Pi$, the language
decided by $M^O$ is consistent with $\Pi'$. Then $\OGTD[\Pi']\subseteq\OGTD[\Pi]$,
$\P^{\Pi'}\subseteq\P^{\Pi}$, and $\FP^{\Pi'}\subseteq\FP^{\Pi}$.
\end{enumerate}
\end{proposition}
Part (b) is folklore (see, e.g., \cite{KT94}), but we include its accurate proof for the sake of self-containedness.
\begin{proof}
(a) Compute the advice and run $D$; for $\Ppoly$, fix any consistent $O$ and hardwire
$G^O(1^n)$. Negating $D$ gives closure under complement.

(b) ``$\subseteq$'': Let $L\in\OGTD[A]$ via $(G,D)$, with $|G^A(1^n)|=p(n)$. Let
$T=\{1^{\pi(n,i)}:1\le i\le p(n),\ \text{bit $i$ of $G^{A}(1^n)$ is $1$}\}$. On input
$1^N$, a polynomial-time machine computes $(n,i)=\pi^{-1}(N)$, checks $1\le i\le p(n)$,
and runs $G^A(1^n)$; since $n\le N$, $T\in\P^A$. On input $x$ of length $n$, a machine
with oracle $T$ queries $1^{\pi(n,i)}$ for $i\le p(n)$, which have polynomial lengths,
and runs $D$. Hence $L\in\P^T$. ``$\supseteq$'': Let $T\in\TALLY\cap\P^A$ and
$L\in\P^T$ via a machine $N$ with running time $q$. Let $G^A(1^n)$ be the string
$T(1^0)T(1^1)\cdots T(1^{q(n)})$, computed with the $\P^A$ machine for $T$, and let
$D(x,s)$ simulate $N$ on $x$, answering a query $w$ by the $(|w|+1)$-st bit of $s$ if
$w=1^{|w|}$, and by $0$ otherwise.

(c) Let $M$ decide $L$ with oracles $L_1,\dots,L_r$ in time $t(n)$, and let
$(G_i,D_i)$ witness $L_i\in\OGTD[\Pi_i]$. Given an oracle $O$ consistent with
$\Pi_1\oplus\dots\oplus\Pi_r$, its restrictions $O_i$ are consistent with $\Pi_i$. The
new advice for length $n$ is the tuple of all $a_{i,m}=G_i^{O_i}(1^m)$ for $i\le r$ and
$m\le t(n)$. The new decoder simulates $M$ and answers a query $q$ to $L_i$ by
$D_i(q,a_{i,|q|})$. For $\P^{\OGTD[\Pi]}=\OGTD[\Pi]$, take $r=1$; the
containment ``$\supseteq$'' is trivial.

(d) Replace every query of the generator for $\Pi'$ by a simulation of $M^O$; the
simulated answers form a language consistent with $\Pi'$. The same works for $\P^{\Pi'}$
and $\FP^{\Pi'}$.
\end{proof}

By part (d), a $\prSBP$-complete problem can replace any promise
problem in $\prSBP$; thus $\OGTD[\SSE]=\bigcup_{\Pi\in\prSBP}\OGTD[\Pi]$, which we
denote $\OGTD[\prSBP]$.
\section{Linear orders and the collapse to \texorpdfstring{$\OMIN$}{OMIN}}\label{sec:orders}

\subsection{Closure under polynomial-time Turing reductions}

The following lemma is the non-oblivious analogue of $\P^{\OMIN}=\OMIN$
\cite[Theorem~4.1]{HVW26}. For a reduction-based definition of $\LtP$ it would
be immediate; for Definition~\ref{def:L2} we prove it for the sake of self-containedness.

\begin{lemma}\label{lem:closure}
$\LtP$ is closed under polynomial-time Turing reductions: if
$A_1,\dots,A_r\in\LtP$, then $\P^{A_1,\dots,A_r}\subseteq\LtP$.
\end{lemma}

\begin{proof}
\emph{Reduction to one oracle.} Let $A=\{\langle i,q\rangle:1\le i\le r,\ q\in
A_i\}$. Then $A\in\LtP$. On input $\langle i,q\rangle$ use the order of $A_i$ on
$q$, padded to a common length, and on all other inputs use the lexicographic order,
whose minimum $0\cdots0$ starts with~$0$. Since
$\P^{A_1,\dots,A_r}\subseteq\P^{A}$, it suffices to treat a single oracle
$A\in\LtP$, given by $V_A$ and a polynomial $p_A$.

\emph{Setup.} Let $L\in\P^A$ via a deterministic oracle machine $M$ whose running
time is bounded by a polynomial $t$. Fix an input $x$ of length $n$, and put
$K=t(n)$ and $P=p_A(t(n))$. For a query $q$ with $|q|\le t(n)$, let $\sqsubset_q$
be the strict linear order on $\bit^P$ that compares the first $p_A(|q|)$ bits by
$\prec_q$ and, in case of equality, the remaining bits lexicographically. Its
minimum starts with $A(q)$. Let $\sqsubset_\bot$ be the lexicographic order on
$\bit^P$. For $c=(c_1,\dots,c_K)\in(\bit^P)^K$, run $M$ on $x$, answering the
$j$-th query by the first bit of $c_j$. Let $q_j(c)$ be the $j$-th query of this
run, or $\bot$ if there are fewer than $j$ queries, and let $\out(c)$ be the output.
Then $q_j(c)$ depends only on $c_1,\dots,c_{j-1}$.

\emph{The order.} For $c\neq c'$, let $j$ be the least index with $c_j\neq c'_j$.
Then $q_j(c)=q_j(c')=:q$, and we define the order $\sqsubset$ on tuples by
\[
 c\sqsubset c'\quad \Longleftrightarrow\quad c_j\sqsubset_q c'_j .
\]
This relation is irreflexive, and any two distinct elements are comparable. For
transitivity, let $c\sqsubset c'\sqsubset c''$, with first differences at $j$ (for
$c,c'$) and at $j'$ (for $c',c''$). If $j<j'$, then $c$ and $c''$ first differ at $j$,
with $c''_j=c'_j$. If $j>j'$, then they first differ at $j'$, with $c_{j'}=c'_{j'}$.
If $j=j'$, then $c_j\sqsubset_q c'_j\sqsubset_q c''_j$, so they first differ at
$j$. In each case the order used at that index is the same, because it is
determined by the common prefix. Hence $c\sqsubset c''$, and $\sqsubset$ is a strict linear
order.

\emph{The minimum.} Define $c^*$ inductively by $c^*_j=\min_{\sqsubset_{q_j(c^*)}}$;
this is well defined because $q_j(c^*)$ depends only on $c^*_1,\dots,c^*_{j-1}$.
Then $c^*$ is the minimum of $\sqsubset$. Indeed, if $c\neq c^*$ first differs from
$c^*$ at $j$, then $c^*_j\sqsubset_q c_j$ for $q=q_j(c)=q_j(c^*)$. By induction on
$j$, the first bit of $c^*_j$ equals $A(q_j(c^*))$ whenever $q_j(c^*)\neq\bot$. So
the run determined by $c^*$ is the true run of $M^A$ on $x$, and
$\out(c^*)=L(x)$.

\emph{The answer bit.} On pairs $(b,c)\in\bit\times(\bit^P)^K$, written as
strings starting with $b$, define
\[
 (b,c)\sqsubset'(b',c')\quad \Longleftrightarrow\quad c\sqsubset c',\ \text{ or }\
 c=c'\ \text{ and }\ b=\out(c)\neq b'.
\]
This is a strict linear order, the pullback of the lexicographic product of $\sqsubset$
and the order $0<1$ under the injection $(b,c)\mapsto(c,b\oplus\out(c))$. Its minimum is
$(L(x),c^*)$, which starts with $L(x)$. It can be decided in time polynomial in
$n$: find the first differing index, simulate $M$ to obtain the query there, and
evaluate $V_A$. Hence $L\in\LtP$, with strings of length $1+t(n)\,p_A(t(n))$.
\end{proof}

\subsection{Tally languages make linear orders oblivious}

\begin{lemma}\label{lem:tally}
If $T$ is a tally language in $\LtP$, then $\P^T\subseteq\OMIN$.
\end{lemma}

\begin{proof}
Let verifier $V$ and polynomial $p$ define $T\in\LtP$. The only input of length $n$ that may be in
a tally language is $1^n$, so $T\in\OMIN$ via the orders $\prec_{1^n}$, which depend only on $n$, and
the decoder that outputs the first bit of the minimum on input $1^n$ and $0$ on all
other inputs. Hence $\P^T\subseteq\P^{\OMIN}=\OMIN$~\cite[Theorem~4.1]{HVW26}.
\end{proof}

\begin{theorem}[part (1) of Theorem~\ref{thm:C}]\label{thm:OPL2}
$(\OLtP=)\;\OMIN=\OGTD[\LtP]=\P^{\TALLY\cap\LtP}$.
\end{theorem}

\begin{proof}
By Proposition~\ref{prop:OP}(b) and Lemma~\ref{lem:closure},
$\OGTD[\LtP]=\bigcup_{A\in\LtP}\P^{\TALLY\cap\P^{A}}=\P^{\TALLY\cap\LtP}$, and
$\P^{\TALLY\cap\LtP}\subseteq\OMIN$ by Lemma~\ref{lem:tally}. It remains to show
$\OMIN\subseteq\OGTD[\LtP]$. Let $L\in\OMIN$ be defined by a polynomial $p$,
decoder $D$, and order $<_n$ with oblivious minima $\mu_n$, and
let
\[
 T=\{1^{\pi(n,i)}:1\le i\le p(n),\ \text{bit $i$ of $\mu_n$ is $1$}\}.
\]
Clearly $L\in\OGTD[T]$: the advice for length $n$ is $\mu_n$, computed with the
queries $1^{\pi(n,i)}$, and the decoder is $D$. We show $T\in\LtP$. Let $y$ be an
input of length $N$. If $y=1^N$ and $\pi^{-1}(N)=(n,i)$ with $1\le i\le p(n)$, the
elements are pairs $(b,u)$ with $b\in\bit$ and $u\in\bit^{p(N)}$. Note that $p(n)\le p(N)$, since
$n\le N$. Pairs are compared first by $u$, comparing the first $p(n)$ bits by
$<_n$ and the remaining bits lexicographically. If $u=u'$, then $(b,u)$ precedes
$(b',u)$ iff $b=u[i]\neq b'$. This is a strict linear order, and its minimum is
$(\mu_n[i],\mu_n0\cdots0)$, which starts with $T(y)$. For all other inputs $y$, use the
lexicographic order on $\bit^{1+p(N)}$, whose minimum starts with $0=T(y)$. The
order is decidable in time polynomial in $N$. Finally, $\OLtP=\OMIN$ by \ref{fact:F2}.
\end{proof}

For the sake of self-containedness,
we provide a direct proof of \ref{fact:F3} from Definition~\ref{def:L2} 
(without the equivalence discussed in Remark~\ref{rem:defs}).

\begin{lemma}[\ref{fact:F3} from order-based $\LtP$]\label{lem:prg}
Fact~\ref{fact:F3} holds for $\LtP$ as in Definition~\ref{def:L2}.
\end{lemma}

\begin{proof}
Korten~\cite{Kor22} gives a deterministic polynomial-time algorithm that computes from
$1^N$ a circuit $C_N$ with more outputs than inputs, and that converts every string
outside the range of $C_N$ into a list $g_1,\dots,g_m$ as in~\ref{fact:F3}
(see~\cite[Proposition~5.5]{HV26}). By Korten's stretch amplification~\cite{Kor22}, we
may assume that $C_N$ has exactly twice as many outputs as inputs: a string outside the
range of the amplified circuit is converted into a string outside the range of $C_N$ in
polynomial time with an $\NP$ oracle. For every circuit $C$ with twice as many outputs
as inputs, Korten and Pitassi~\cite[Theorem~1]{KP24} (see~\cite[Section~3]{HVW26}) give
a strict linear order $<_C$, decidable in polynomial time, and a polynomial-time
algorithm that converts $\min{<_C}$ into a string outside the range of $C$. Let $A$ be
the language of pairs $(C,i)$ such that bit $i$ of $\min{<_C}$ is $1$. Then
$A\in\LtP$: on input $(C,i)$ with $1\le i\le m$, where $\bit^m$ is the domain of $<_C$,
compare the first $m$ bits of two strings by $<_C$ after moving the first bit to
position $i$, and the remaining bits lexicographically; on all other inputs use the
lexicographic order. Hence the list for
$1^N$ is computed in deterministic polynomial time with the oracles $A$ and $\SAT$, and
the tally language $T=\{1^{\pi(N,j)}:\text{bit $j$ of the list for $1^N$ is }1\}$ is
in $\P^{A,\SAT}$. By~\ref{fact:F1} and Lemma~\ref{lem:closure}, $T\in\LtP$, and by
Lemma~\ref{lem:tally}, $T\in\P^T\subseteq\OMIN=\OLtP$. Thus $B=T$ satisfies~\ref{fact:F3}.
\end{proof}

\subsection{Replacing Merlin--Arthur oracles by oblivious linear orders}

Theorem~\ref{thm:OPprMA} is the input-oblivious counterpart of
$\P^{\prMA}\subseteq\LtP$~\cite{HV26}; it does not follow from the stronger
containment $\P^{\prMA}\subseteq\P^{\OLtP,\NP}$~\cite[Corollary~5.6]{HV26} (see
also~\cite[Proposition~6.9]{HVW26}), because $\NP\subseteq\OLtP$ is equivalent to
$\NP\subseteq\Ppoly$ (Corollary~\ref{cor:equiv}), but it follows easily from the proof of
the latter and Theorem~\ref{thm:OPL2}.

\begin{theorem}[part (2) of Theorem~\ref{thm:C}]\label{thm:OPprMA}
$\OGTD[\prMA]\subseteq\OLtP$.
\end{theorem}

\begin{proof}
Let polynomial $s$ and verifier $V$ define $\Pi\in\prMA$, with perfect completeness. Let $B\in\OLtP$ be the
language from \ref{fact:F3}. We describe a deterministic polynomial-time oracle machine $M$
with oracle $B\oplus\SAT$ such that the language $O'$ decided by $M^{B\oplus\SAT}$ is
consistent with $\Pi$. On a query $q$ of length $\ell$, let $S(\ell)$ be a polynomial
bound on the size of the circuits $r\mapsto V(q,w,r)$ for $|q|=\ell$ and
$|w|=s(\ell)$, and put $N=\max(S(\ell),s(\ell),4)$. The machine $M$ computes the list
$g_1,\dots,g_m\in\bit^N$ of \ref{fact:F3} with oracle $B$. It lets $g'_i$ be the
$s(\ell)$-bit prefix of $g_i$, and accepts iff
\[
 \exists w\in\bit^{s(\ell)}\ \ \forall i\le m\ \ V(q,w,g'_i)=1,
\]
which it decides with one $\SAT$ query. If $q\in\Pi_\YES$, the witness $w$ from
perfect completeness passes all tests, so $q\in O'$. If $q\in\Pi_\NO$, then for
every $w$, the circuit that computes $V(q,w,\cdot)$ on the first $s(\ell)$ of its
$N$ inputs has size at most $N$. Hence
$\Pr_i[V(q,w,g'_i)=1]<1/4+1/N<1$, so some test fails and $q\notin O'$.

By Proposition~\ref{prop:OP}(d), applied to the language $B\oplus\SAT$ (which is
consistent only with itself), $\OGTD[\Pi]\subseteq\OGTD[B\oplus\SAT]$. Now
$B\oplus\SAT\in\LtP$ by \ref{fact:F1}, \ref{fact:F2}, and Lemma~\ref{lem:closure}. Hence
$\OGTD[\Pi]\subseteq\OGTD[\LtP]=\OLtP$ by Theorem~\ref{thm:OPL2}.
\end{proof}

\begin{remark}[input-oblivious oracles]\label{rem:monotone}
The class $\OGTD[\mathcal X]$ is monotone in $\mathcal X$; in particular,\linebreak%
$\OGTD[\prOMA]\subseteq\OGTD[\prMA]$. For generators that make non-adaptive queries the
two classes coincide. Indeed, if the $i$-th query $q_i(n)$ is computable from $1^n$,
the generator can ask instead the tally instance $1^{\pi(n,i)}$, which is a
yes-instance iff $q_i(n)\in\Pi_\YES$ and a no-instance iff $q_i(n)\in\Pi_\NO$; and a
promise problem with at most one instance of each length is in $\prOMA$ as soon as it
is in $\prMA$. We do not know whether $\OGTD[\prOMA]=\OGTD[\prMA]$ in general. If
$\NP\subseteq\Ppoly$, both classes equal $\PH$ (Corollary~\ref{cor:main-KL}, and
Theorem~\ref{thm:transfer} with~\ref{fact:F4}).
\end{remark}

\begin{remark}
The same machine shows $\P^{\prMA}\subseteq\P^{B\oplus\SAT}\subseteq\LtP$ for
Definition~\ref{def:L2}; this is the argument of~\cite[Theorem~5.4 and
Corollary~5.6]{HV26}, with Lemma~\ref{lem:closure} supplying the closure step. The
point of Theorem~\ref{thm:OPprMA} is that after the replacement, the oracle is a
language. Therefore the advice computed from $1^n$ is unique, and the tally argument
of Theorem~\ref{thm:OPL2} applies.
\end{remark}

\subsection{The collapse to \texorpdfstring{$\OMIN$}{OMIN}}\label{sec:KL}

\begin{lemma}[folklore, cf.~\cite{KL80}]\label{lem:ppoly}
If $\NP\subseteq\Ppoly$, then $\PH\subseteq\Ppoly$.
\end{lemma}

\begin{proof}
We show $\SigmakP{k}\subseteq\Ppoly$ by induction on $k$; the case $k=1$ is the
hypothesis. Let $L\in\SigmakP{k+1}$, so $L$ is accepted by a nondeterministic oracle
machine $N$ with an oracle $A\in\SigmakP{k}$ and running time bounded by a polynomial
$t$. By induction, $A\cap\bit^m$ has circuits $A_m$ of size at most $a(m)$, for a
polynomial $a$. Let $L'$ be the set of pairs $(x,z)$ such that $z$ encodes circuits
$E_0,\dots,E_{t(|x|)}$, with $E_m$ of size at most $a(m)$, and $N$ has an accepting
computation on $x$ in which every query $w$ is answered by $E_{|w|}(w)$. Then
$L'\in\NP$, and $x\in L$ iff $(x,z_{|x|})\in L'$, where $z_n$ encodes
$A_0,\dots,A_{t(n)}$. The length of $z_n$ is polynomial and depends only on $n$.
Since $L'\in\NP\subseteq\Ppoly$, hardwiring $z_n$ into the circuits for $L'$
gives polynomial-size circuits for $L$.
\end{proof}

\begin{lemma}[canonical circuits]\label{lem:canonical}
If $\NP\subseteq\Ppoly$, then for every $L\in\PH$ there is a tally language
$T_L\in\PH$ such that $L\in\P^{T_L}$. More precisely, there are a polynomial
$\ell$ and a polynomial-time oracle machine that, on input $1^n$, queries
$1^{\pi(n,1)},\dots,1^{\pi(n,\ell(n))}$ and outputs the lexicographically first
circuit of a fixed polynomial size computing $L$ on $\bit^n$.
\end{lemma}

\begin{proof}
Let $L\in\SigmakP{k}$, $k\ge1$. By Lemma~\ref{lem:ppoly}, there is a constant $c$
such that $L\cap\bit^n$ is computed by a circuit with $n$ inputs and at most
$s(n)=n^c+c$ gates, for every $n$. Put $\ell(n)=\ell(n,s(n))$. Let $C_n$ be the
lexicographically first string of length $\ell(n)$ that encodes a circuit computing
$L$ on $\bit^n$, and define
\[
 T_L=\bigl\{1^{\pi(n,i)}:\ n\ge0,\ 1\le i\le\ell(n),\ \text{the $i$-th bit of
 $C_n$ is $1$}\bigr\}.
\]
On input $x\in\bit^n$, query $1^{\pi(n,i)}$ for $i=1,\dots,\ell(n)$, which yields
$C_n$, and output $C_n(x)$; so $L\in\P^{T_L}$. To show $T_L\in\PH$, let $Q$ be the
set of pairs $(1^n,\alpha)$ such that some $w\in\bit^{\ell(n)}$ with prefix $\alpha$
encodes a circuit $C_w$ with $\forall x\in\bit^n\ C_w(x)=L(x)$. The predicate
$C_w(x)=L(x)$ is in $\P^{\SigmakP{k}}$, so $Q\in\SigmakP{k+2}$. To decide $1^N$,
compute $(n,i)=\pi^{-1}(N)$ and reject unless $1\le i\le\ell(n)$. Otherwise compute
$C_n$ bit by bit with $\ell(n)$ queries to $Q$: if $\alpha$ is the prefix found so
far, the next bit is $0$ iff $(1^n,\alpha0)\in Q$. Accept iff the $i$-th bit of
$C_n$ is $1$. This takes time polynomial in $N$, so $T_L\in\P^{\SigmakP{k+2}}\subseteq\PH$.
\end{proof}

\begin{theorem}[transfer principle]\label{thm:transfer}
Let $\mathcal C$ be a class of promise problems (possibly languages) such that
$\NP\subseteq\Ppoly$ implies $\PH\subseteq\P^{\mathcal C}$. Then
$\NP\subseteq\Ppoly$ implies $\PH\subseteq\OGTD[\mathcal C]$.
\end{theorem}

\begin{proof}
Assume $\NP\subseteq\Ppoly$ and let $L\in\PH$. By Lemma~\ref{lem:canonical},
$L\in\P^{T_L}$ for a tally language $T_L\in\PH$. By hypothesis, $T_L\in\P^{\Pi}$
for some $\Pi\in\mathcal C$. A tally language in $\P^\Pi$ is in $\OGTD[\Pi]$: the advice
for length $N$ is the bit $T_L(1^N)$, computed by the $\P^\Pi$ machine for $T_L$, which is correct with every consistent oracle.
The decoder outputs this bit on $1^N$ and $0$ on all other inputs. Hence
$L\in\P^{T_L}\subseteq\P^{\OGTD[\Pi]}=\OGTD[\Pi]$ by Proposition~\ref{prop:OP}(c).
\end{proof}

\begin{corollary}[Theorem~\ref{thm:ominKL}]\label{cor:main-KL}
If $\NP\subseteq\Ppoly$, then
\[
 \PH=\OGTD[\prMA]=\OLtP=\OMIN .
\]
Consequently, every class $\mathcal D$ with $\OGTD[\prMA]\subseteq\mathcal
D\subseteq\PH$ equals $\PH$. This includes all classes of Figure~\ref{fig:mirror}.
\end{corollary}

\begin{proof}
By~\cite{CR11}, $\NP\subseteq\Ppoly$ implies $\PH=\P^{\prMA}$, so
$\PH\subseteq\OGTD[\prMA]$ by Theorem~\ref{thm:transfer}. By
Theorem~\ref{thm:OPprMA}, $\OGTD[\prMA]\subseteq\OLtP$, and $\OLtP=\OMIN\subseteq\PH$.
\end{proof}

\begin{remark}%
\label{rem:routes}
Appendix~\ref{app:routes} gives two other routes to the collapse to $\OMIN$: through
Theorem~\ref{thm:B} instead of the collapse of~\cite{CR11}, and through the collapse to
$\ZPP^{\NP}$~\cite{KW98}, which avoids promise problems. It also discusses
relativization.
\end{remark}

\begin{corollary}\label{cor:equiv}
The following are equivalent:
\textup{(i)} $\NP\subseteq\Ppoly$;
\textup{(ii)} $\NP\subseteq\OLtP$;
\textup{(iii)} $\PH=\OLtP$;
\textup{(iv)} $\PH\subseteq\OGTD[\prMA]$;
\textup{(v)} $\P^{\prSBP}\subseteq\Ppoly$.
\end{corollary}

\begin{proof}
(i)$\Rightarrow$(iv) and (i)$\Rightarrow$(iii) by Corollary~\ref{cor:main-KL};
(iv)$\Rightarrow$(iii) since $\OGTD[\prMA]\subseteq\OLtP\subseteq\PH$;
(iii)$\Rightarrow$(ii) is trivial; and (ii)$\Rightarrow$(i) since
$\OLtP\subseteq\Ppoly$. Finally, (i) implies $\P^{\prSBP}\subseteq\PH=\OLtP\subseteq\Ppoly$,
and (v) implies (i) since $\NP\subseteq\P^{\prSBP}$.
\end{proof}

\begin{corollary}[minimality]\label{cor:minimal}
$\OLtP$ has a Karp--Lipton collapse, and $\OLtP\not\subseteq\SIZE[n^k]$ for every
$k$. It is contained in $\OtP\cap\LtP$, and no smaller class with both properties is
known. Section~\ref{sec:ultimate} gives a collapse to the smaller class $\OM$.
\end{corollary}

\begin{proof}
By Corollary~\ref{cor:main-KL}, \ref{fact:F2}, \ref{fact:F4}, and Proposition~\ref{prop:OP}(a).
\end{proof}

\section{The ultimate collapse}\label{sec:ultimate}

\subsection{Strict oblivious Merlin--Arthur oracles}

The pointwise domination condition \eqref{cond:OMA4} lets us sample an $\sOprMA$ oracle with an
$\ONP$ oracle. The following lemma is stated for transducers, because we apply it to
generators.

\begin{lemma}[sampling $\sOprMA$ oracles]\label{lem:sample}
For every $\Pi\in\sOprMA$ there is a language $K_\Pi\in\ONP$ with the following
property. For every deterministic polynomial-time oracle transducer $M$ there is a
randomized polynomial-time oracle transducer $M'$ such that, for every input $z$, with
probability at least $3/4$, $M'^{K_\Pi}(z)=M^O(z)$ for some language $O$ consistent
with $\Pi$. In particular, $\P^{\sOprMA}\subseteq\BPP^{\ONP}$.
\end{lemma}

\begin{proof}
Let $V$ and $(y_n)$ witness $\Pi\in\sOprMA$, and let
$K_\Pi=\{(x,r):\exists y\ V(x,y,r)=1\}$. By \eqref{cond:OMA4}, $(x,r)\in K_\Pi$ iff
$V(x,y_{|x|},r)=1$, so $y_{|x|}$ is an oblivious witness and $K_\Pi\in\ONP$. If
$x\in\Pi_\YES$, then $(x,r)\in K_\Pi$ for every $r$. If $x\in\Pi_\NO$, then
$\Pr_r[(x,r)\in K_\Pi]=\Pr_r[V(x,y_{|x|},r)=1]<1/4$.

Let $Q\ge1$ bound the number of queries of $M$. The transducer $M'$ simulates $M$. At the
first occurrence of a query $x$, it chooses independent random strings $r_1,\dots,r_k$,
where $4^{-k}\le1/(4Q)$, and answers YES iff $(x,r_j)\in K_\Pi$ for all $j$. It stores
the answer and reuses it for later occurrences of $x$. Queries in $\Pi_\YES$ are always
answered YES, and each query in $\Pi_\NO$ is answered YES with probability less than
$1/(4Q)$. By the union bound, with probability at least $3/4$ all queries in
$\Pi_\YES\cup\Pi_\NO$ are answered correctly. The stored answers then extend to a
language $O$ consistent with $\Pi$, and $M'^{K_\Pi}(z)=M^O(z)$.
\end{proof}

The following lemma is implicit in the proof of~\cite[Proposition~6.9]{HVW26}, which
establishes the collapse to $\P^{\OprMA}$.

\begin{lemma}[collapse to $\P^{\sOprMA}$]\label{lem:soprmaKL}
If $\NP\subseteq\Ppoly$, then every promise problem in $\prMA$ belongs to $\sOprMA$.
Consequently, $\PH=\P^{\sOprMA}$.
\end{lemma}

\begin{proof}
Let $\Pi\in\prMA$ via $V(x,y,r)$ with perfect completeness, soundness $1/4$, and proofs
of length $p(n)$. Let $t=p(n)+2$, and for $R=(r_1,\dots,r_t)$ let
$C_{x,R}(y)=\bigwedge_{i\le t}V(x,y,r_i)$. If $x\in\Pi_\NO$, then
\[
 \Pr_R[\exists y\ C_{x,R}(y)=1]\le2^{p(n)}4^{-t}<1/4 .
\]
Since $\NP\subseteq\Ppoly$, search-to-decision gives polynomial-size circuits $E_n$ that
output a satisfying assignment of every satisfiable circuit of the size of $C_{x,R}$,
for $x\in\bit^n$. The new verifier $V'(x,E,R)$ rejects unless $E$ encodes a circuit of
the prescribed size; otherwise it computes $y=E(C_{x,R})$ and accepts iff
$C_{x,R}(y)=1$. The oblivious proof is $E_n$.
\begin{itemize}
\item[\eqref{cond:OMA1}] If $x\in\Pi_\YES$, then $C_{x,R}$ is satisfiable for every $R$, so
$V'(x,E_n,R)=1$.
\item[\eqref{cond:OMA2}] If $x\in\Pi_\NO$, then $V'(x,E,R)=1$ implies that $C_{x,R}$ is
satisfiable, which happens with probability less than $1/4$.
\item[\eqref{cond:OMA4}] For every $x$, $E$ and $R$, if $V'(x,E,R)=1$, then $C_{x,R}$ is
satisfiable, so $V'(x,E_n,R)=1$.
\end{itemize}
Finally, $\PH=\P^{\prMA}$~\cite{CR11}, $\P^{\prMA}\subseteq\P^{\sOprMA}$ under the
hypothesis, and $\P^{\sOprMA}\subseteq\P^{\prMA}\subseteq\PH$.
\end{proof}

\begin{observation}[collapse to $\ZPP^{\ONP}$]\label{obs:zpponp}
If $\NP\subseteq\Ppoly$, then $\NP=\ONP$ and $\PH=\ZPP^{\ONP}$.
\end{observation}

\begin{proof}
Let $L\in\NP$ via a verifier $U(x,w)$. By the hypothesis and search-to-decision, there
are polynomial-size circuits $S_n$ that output a witness for every $x\in L\cap\bit^n$.
The verifier $U'(x,S)=U(x,S(x))$ has the oblivious witness $S_n$ and never accepts
$x\notin L$. Hence $\NP=\ONP$, which is~\cite[Proposition~5]{FSW09}, and
$\PH=\ZPP^{\NP}=\ZPP^{\ONP}$~\cite{KW98}.
\end{proof}

Lemma~\ref{lem:soprmaKL}, Observation~\ref{obs:zpponp} and Theorem~\ref{thm:ominKL}
are the three deepest collapses discussed in the introduction.

\subsection{The class \texorpdfstring{$\OM$}{OM}}

Recall that $\OM=\OGTD[\FP^{\sOprMA},\co\ONP,\P]$ and
$\OUM=\OUGTD[\FP^{\sOprMA},\co\ONP,\P]$. Like $\OMIN$, these classes are ``robust'',
i.e., closed under complement and polynomial-time Turing reductions.

\begin{proposition}\label{prop:OMclosure}
$\OM$ and $\OUM$ are closed under complement and under polynomial-time Turing
reductions: $\P^{\OM}=\OM$ and $\P^{\OUM}=\OUM$.
\end{proposition}

\begin{proof}
For complement, negate the decoder. Let $L\in\P^{L_1,\dots,L_r}$ via a machine $M$
running in time $t(n)$, where $L_i\in\OM$ via $(G_i,T_i,D_i)$ with oracles
$\Pi_i\in\sOprMA$. The problem $\Pi_1\oplus\dots\oplus\Pi_r$ is in $\sOprMA$: on
input $\langle i,x\rangle$ run the verifier of $\Pi_i$, and use the tuple of the
oblivious proofs as the oblivious proof; \eqref{cond:OMA4} holds componentwise. The new generator
outputs the tuple of all $G_i(1^m)$ for $i\le r$ and $m\le t(n)$, computed in one
run with this oracle. Each component of a legal output is a legal output of $G_i$. The
new tester accepts a tuple iff every component $a_{i,m}$ satisfies
$(1^m,a_{i,m})\in T_i$. Its complement is in $\ONP$, with the tuple of the oblivious
witnesses of the complements of the $T_i$ at all lengths $m\le t(n)$ as the oblivious
witness. The new decoder simulates $M$ and answers a query $q$ to $L_i$ by
$D_i(q,a_{i,|q|})$. If every $T_i$ accepts exactly one string per length, so does the
new tester.
\end{proof}

\subsection{Three ways to run the generator}

\begin{theorem}\label{thm:three}
$\OUM\subseteq\OM\subseteq\OMIN\cap\P^{\sOprMA}\cap\ZPP^{\ONP}$.
\end{theorem}

\begin{proof}
Let $L\in\OM$ via a generator $G$ with oracle $\Pi\in\sOprMA$, a tester
$T\in\co\ONP$, and a decoder $D\in\P$. Let $\overline T\in\ONP$ be the complement of
$T$; pairs not of the form $(1^n,a)$ with $|a|=p(n)$ belong to $\overline T$ and are
accepted without a witness.

\emph{$\P^{\sOprMA}$: run the generator.} On input $x\in\bit^n$, compute
$a=G^O(1^n)$ and output $D(x,a)$. Since $a\in\Out_G(n)$ is good advice, this is
correct for every $O$ consistent with $\Pi$.

\emph{$\OMIN$: derandomize the generator.} Dropping the tester,
$\OM\subseteq\OGTD[\sOprMA]\subseteq\OGTD[\prMA]$, because $\sOprMA\subseteq\prMA$.
By Theorem~\ref{thm:OPprMA}, $\OGTD[\prMA]\subseteq\OLtP=\OMIN$.

\emph{$\ZPP^{\ONP}$: sample the generator and test.} On input $x\in\bit^n$, run the
randomized simulation of $G$ on $1^n$ from Lemma~\ref{lem:sample} with the oracle
$K_\Pi\in\ONP$, obtaining a string $a$. Then query $(1^n,a)\in\overline T$. If the
answer is yes, output ``?''. Otherwise $T$ accepts $a$, so $a$ is good advice, and the
algorithm outputs $D(x,a)=L(x)$. With probability at least $3/4$, the string $a$ is
in $\Out_G(n)$, which $T$ accepts, so the output is not ``?''. The two $\ONP$ oracles
are combined into one $\ONP$ language by tagging inputs and bundling oblivious
witnesses. Hence $L\in\ZPP^{\ONP}$.
\end{proof}

\subsection{The collapse}

\begin{theorem}\label{thm:OMcollapse}
If $\NP\subseteq\Ppoly$, then $\PH=\OUM=\OM$.
\end{theorem}

\begin{proof}
By Theorem~\ref{thm:three}, $\OUM\subseteq\OM\subseteq\OMIN\subseteq\PH$. Conversely,
assume $\NP\subseteq\Ppoly$ and let $L\in\PH$. By Theorem~\ref{thm:ominKL}, $L\in\OMIN$
via an oblivious order $<_n$ on $\bit^{p(n)}$, minima $\mu_n$, and a decoder $D$.
\begin{itemize}
\item \emph{Tester.} Let $T=\{(1^n,a):a\in\bit^{p(n)}\text{ and no }u\text{ satisfies }
u<_n a\}$. Every $a\ne\mu_n$ satisfies $\mu_n<_n a$, so the complement of $T$ is in
$\ONP$ with the oblivious witness $\mu_n$. Thus $T\in\co\ONP$, and $T$ accepts exactly
$\mu_n$ at length $n$.
\item \emph{Decoder.} $D$ itself: $D(x,\mu_n)=L(x)$ for every $x\in\bit^n$.
\item \emph{Generator.} The tally language
$M=\{1^{\pi(n,i)}:1\le i\le p(n),\ \text{bit $i$ of $\mu_n$ is $1$}\}$ is in $\LtP$
(see the proof of Theorem~\ref{thm:OPL2}), hence in $\PH$. By
Lemma~\ref{lem:soprmaKL}, $M\in\P^{\Pi}$ for some $\Pi\in\sOprMA$. The generator
computes $\mu_n$ bit by bit with this $\P^{\Pi}$ machine. Every consistent oracle gives
the correct bits, so $\Out_G(n)=\{\mu_n\}$.
\end{itemize}
Hence \eqref{eq:OGTD} holds with $\mathrm{Acc}_T(n)=\{\mu_n\}$, and $L\in\OUM$.
\end{proof}

\begin{remark}\label{rem:OM}
\begin{enumerate}
\item The proof combines two of the three collapses: $\OMIN$ supplies a canonical
advice string with a $\co\ONP$ certificate, and $\P^{\sOprMA}$ supplies the generator. The
certificate is the one behind $\OMIN\subseteq\SOUP^{\ONP[1]}$~\cite[Theorem~5.2]{HVW26},
which, combined with Theorem~\ref{thm:three}, also gives $\OM\subseteq\SOUP^{\ONP[1]}$.
\item The collapse to $\OM$ can also be proved without $\OMIN$, along the lines of the
proof of the collapse to $\ZPP^{\NP}$~\cite{BCG+96,KW98} (Appendix~\ref{app:learning}).
\item By Theorem~\ref{thm:transfer}, the tally closure $\P^{\TALLY\cap\mathcal C}$ of every Karp--Lipton class $\mathcal{C}$ that is
closed under polynomial-time Turing reductions is again a Karp--Lipton class. So some
Karp--Lipton class inside $\OMIN\cap\P^{\sOprMA}\cap\ZPP^{\ONP}$ is easy to obtain:
the tally closure of $\P^{\sOprMA}\cap\ZPP^{\ONP}$, which lies in $\OMIN$ by
Theorem~\ref{thm:OPL2}, because $\P^{\sOprMA}\subseteq\P^{\prMA}\subseteq\LtP$~\cite{HV26}.
What distinguishes $\OM$ is that it is defined by a natural model, and that each of
the three containments comes from a different way of running its generator. The zero-error simulation
uses the tester: we do not know whether $\OGTD[\sOprMA]\subseteq\ZPP^{\ONP}$.
\item Fixed-polynomial lower bounds for $\OM$ remain open (see discussion in Section~\ref{sec:open}).
\end{enumerate}
\end{remark}
\section{A Goldreich--Wigderson lemma with approximate counting, and \texorpdfstring{$\P^{\prStP}\subseteq\P^{\prSBP}$}{P\textasciicircum prS2P in P\textasciicircum prSBP}}\label{sec:gw}

Throughout this section, the oracle is a fixed language $O$ consistent with $\SSE$,
and ``computable'' means computable in polynomial time with oracle $O$. All
statements hold for every such $O$.

\paragraph{Matrices~\cite[Def.~4.6]{CR11}.}
Let $V:\bit^a\times\bit^b\to\bit$ be a circuit. A row $y$ \emph{beats} a column
$z$ if $V(y,z)=1$, and $z$ beats $y$ if $V(y,z)=0$. A row is \emph{irrefutable} if it
beats every column, and similarly for columns. A set $R$ of rows is a
\emph{row-side collectively irrefutable certificate} if every column is beaten by
some $y\in R$. A set $T$ of columns is a \emph{column-side} one if every row is
beaten by some $z\in T$.

\begin{lemma}[above-average rows]\label{lem:avg}
Let a nonempty set $Y\subseteq\bit^a$, a set $Z\subseteq\bit^b$, and a predicate
$V$ be given by circuits, and let $\eps\in(0,1]$ be rational. For $y\in Y$, let
$\dg(y)=|\{z\in Z:V(y,z)=1\}|$. One can compute, in time polynomial in the circuit sizes
and $1/\eps$, a row $y\in Y$ with
\[
 \dg(y)\ \ge\ \frac{1}{(1+\eps)^{2a}}\cdot\frac{1}{|Y|}\sum_{y'\in Y}\dg(y').
\]
If $Z$ is nonempty, the same holds for columns $z\in Z$, with $\dg(z)=|\{y\in Y:V(y,z)=0\}|$ and $2b$ in
place of $2a$.
\end{lemma}

\begin{proof}
This is the procedure of~\cite[Lemma~3.12]{HV26}, with the average order rank
replaced by the average degree and ``smaller'' replaced by ``larger''. We fix the
bits of $y$ one at a time, maintaining a prefix $\beta$ such that
$Y_\beta=\{y\in Y: y\text{ starts with }\beta\}$ is nonempty; for a set $H$ of
rows, let $\mathrm{av}(H)$ denote the average degree of its elements. In each step,
let $H_0=Y_{\beta0}$ and $H_1=Y_{\beta1}$.
\begin{itemize}
\item If one of $H_0,H_1$ is empty (detected exactly by Lemma~\ref{lem:count}),
append the bit $c$, for which $H_c \neq \varnothing$, to $\beta$.
\item Otherwise, for $c\in\bit$ compute by Lemma~\ref{lem:count} with accuracy
$\eps$ approximations $t_A^c$ of $|H_c|$ and $t_B^c$ of
$|\{(y,z)\in H_c\times Z:V(y,z)=1\}|$, and let $e_c=t_B^c/t_A^c$. Extend $\beta$ by
the bit $c$ with the larger $e_c$.
\end{itemize}
Each estimate satisfies $\mathrm{av}(H_c)/(1+\eps)\le e_c\le(1+\eps)\mathrm{av}(H_c)$.
Since $\mathrm{av}(Y_\beta)$ is a weighted mean of $\mathrm{av}(H_0)$ and
$\mathrm{av}(H_1)$, it is at most their maximum. If the chosen half has the larger
average, its average is at least $\mathrm{av}(Y_\beta)$. Otherwise its average is
at least $e_c/(1+\eps)\ge e_{1-c}/(1+\eps)\ge\mathrm{av}(H_{1-c})/(1+\eps)^2\ge
\mathrm{av}(Y_\beta)/(1+\eps)^2$. So each step loses a factor of at most
$(1+\eps)^2$, and after $a$ steps $Y_\beta$ consists of a single element $y$ with the
claimed degree.
\end{proof}

\begin{theorem}[Goldreich--Wigderson lemma with approximate counting]\label{thm:gw}
Given any circuit $V:\bit^a\times\bit^b\to\bit$, one can compute a row-side or a
column-side collectively irrefutable certificate with at most $3(a+b)+1$ elements.
\end{theorem}

\begin{proof}
Let $d=\max(a,b,1)$. Maintain lists $R$ and $T$, initially empty, and the sets
\[
 Y=\{y:\forall z\in T\ V(y,z)=1\},\qquad Z=\{z:\forall y\in R\ V(y,z)=0\}
\]
of rows and columns not yet beaten. Both are given by conjunctions of copies of
$V$. Repeat the following steps.
\begin{enumerate}
\item If $Z=\emptyset$, output $R$. If $Y=\emptyset$, output $T$.
Lemma~\ref{lem:count} tests emptiness exactly.
\item Let $N_c=|\{(y,z)\in Y\times Z:V(y,z)=c\}|$ for $c\in\bit$. Compute estimates
$t_0,t_1$ with accuracy~$1$, and let $c=1$ if $t_1\ge t_0$, and $c=0$ otherwise.
Then $N_{1-c}\le t_{1-c}\le t_c\le2N_c$, so $N_c\ge|Y||Z|/3$.
\item If $c=1$, apply Lemma~\ref{lem:avg} with $\eps=1/(8d)$, so that
$(1+\eps)^{2a}\le e^{1/4}<4/3$. It gives $y\in Y$ that beats more than $|Z|/4$
columns of $Z$, and we append $y$ to $R$. If $c=0$, we symmetrically append to $T$
a column of $Z$ that beats more than $|Y|/4$ rows of $Y$.
\end{enumerate}
Each round multiplies $|Y|\cdot|Z|$ by less than $3/4$. This product starts at
$2^{a+b}$ and is at least $1$ while the procedure runs, so there are at most
$3(a+b)+1$ rounds. The output is a certificate by the definitions of $Y$ and $Z$.
\end{proof}

\begin{corollary}\label{cor:irrefutable}
If $V$ has an irrefutable row, the procedure of Theorem~\ref{thm:gw} outputs a
row-side collectively irrefutable certificate $R$ with $|R|\le3(a+b)+1$, for every
oracle consistent with $\SSE$. Symmetrically for columns.
\end{corollary}

\begin{proof}
An irrefutable row is never beaten by a column, so it stays in $Y$, and the
procedure never outputs $T$. Since the number of rounds is bounded, it outputs $R$.
\end{proof}

\begin{theorem}[Theorem~\ref{thm:B}]\label{thm:prS2}
For every $\Pi\in\prStP$ there is a deterministic polynomial-time oracle machine
$G$ such that, for every language $O$ consistent with $\SSE$, the language decided
by $G^O$ is consistent with $\Pi$. Consequently, $\P^{\prStP}\subseteq\P^{\prSBP}$,
$\FP^{\prStP}\subseteq\FP^{\prSBP}$ and $\OGTD[\prStP]\subseteq\OGTD[\prSBP]$. In
particular, $\StP\subseteq\P^{\prSBP}$.
\end{theorem}

\begin{proof}
Let $\Pi\in\prStP$ via $V$. On a query $q$, the machine $G$ runs the procedure of
Theorem~\ref{thm:gw} on $V_q(y,z)=V(q,y,z)$ and answers YES iff the output is
row-side. If $q\in\Pi_\YES$, $V_q$ has an irrefutable row, so the answer is YES by
Corollary~\ref{cor:irrefutable}; symmetrically, if $q\in\Pi_\NO$, the answer is NO.
The remaining claims follow from Proposition~\ref{prop:OP}(d).
\end{proof}

\section{The oblivious setting of symmetric alternation}\label{sec:mirror}

We now prove part~(3) of Theorem~\ref{thm:C}. The basic tool is a list of
witnesses that collectively certify all yes-instances of a given length.

\begin{lemma}[collective oblivious witnesses]\label{lem:collective}
Let $W(x,w)$ be a predicate computable in time polynomial in $|x|$, where $|w|=p(|x|)$
for a polynomial $p$, and possibly depending on an auxiliary string $\alpha$ that is
part of the input of the procedure below. Let $n$ be such that some $w^*$ satisfies
\begin{equation}\label{eq:dominating}
 \forall x\in\bit^n:\quad \bigl(\exists w\ W(x,w)=1\bigr)\ \Rightarrow\ W(x,w^*)=1.
\end{equation}
Then, given $1^n$ (and $\alpha$), one can compute in $\FP^{\SSE}$ a list
$w_1,\dots,w_k$ with $k\le3(2p(n)+n)+1$ such that
\[
 \forall x\in\bit^n:\quad \bigl(\exists w\ W(x,w)=1\bigr)\ \Rightarrow\ \exists j\le k\
 W(x,w_j)=1,
\]
for every oracle consistent with $\SSE$.
\end{lemma}

\begin{proof}
Consider the \emph{domination game} with rows $w\in\bit^{p(n)}$ and columns
$(x,w')\in\bit^{n}\times\bit^{p(n)}$, where row $w$ beats column $(x,w')$ iff
$W(x,w)=1$ or $W(x,w')=0$. It is given by a circuit of size polynomial in $n$ (and
$|\alpha|$). By~\eqref{eq:dominating}, $w^*$ is an irrefutable row. By
Corollary~\ref{cor:irrefutable}, we obtain a list $R=\{w_1,\dots,w_k\}$ such that
every column is beaten by some $w_j$. If $W(x,w')=1$ for some $w'$, then the
column $(x,w')$ is beaten by some $w_j$, which means $W(x,w_j)=1$.
\end{proof}

\begin{theorem}\label{thm:ONP}
$\ONP\subseteq\OGTD[\prSBP]$.
\end{theorem}

\begin{proof}
Let $L\in\ONP$ via a verifier $V(x,w)$ with oblivious witnesses $w_n$. Then $W=V$
satisfies~\eqref{eq:dominating} with $w^*=w_n$: if $V(x,w)=1$ for some $w$, then
$x\in L$, so $V(x,w_n)=1$. The advice for length $n$ is the list of
Lemma~\ref{lem:collective}, padded to a fixed polynomial length by repeating $w_1$,
and the decoder accepts $x$ iff $V(x,w_j)=1$ for some
$j$. If $x\in L$, some $w_j$ works; if $x\notin L$, no witness works.
\end{proof}

\begin{theorem}[part (3) of Theorem~\ref{thm:C}]\label{thm:O2}
$\OtP\subseteq\OGTD[\prSBP]$.
\end{theorem}

\begin{proof}
Let $L\in\OtP$ via a verifier $V(x,y,z)$ with $|y|=|z|=p(|x|)$ and oblivious
irrefutable certificates $y_n,z_n$:
\[
 \forall x\in L\cap\bit^n\ \forall z\ V(x,y_n,z)=1,\qquad
 \forall x\in\bit^n\setminus L\ \forall y\ V(x,y,z_n)=0 .
\]
Fix $n$ and an oracle consistent with $\SSE$. The advice is computed in two stages.

\emph{Stage 1: the consistency game.} Its rows are pairs
$(y,z)\in\bit^{2p(n)}$, its columns are triples $(x,z',y')\in\bit^{n+2p(n)}$.
Let us say that
\[
 \text{row }(y,z)\text{ beats column }(x,z',y')\quad \Longleftrightarrow\quad
 V(x,y,z')=1\ \text{ or }\ V(x,y',z)=0 .
\]
The row $(y_n,z_n)$ is irrefutable: if $x\in L$, then $V(x,y_n,z')=1$, and if
$x\notin L$, then $V(x,y',z_n)=0$. By Corollary~\ref{cor:irrefutable} we compute a
list $R=\{(y_1,z_1),\dots,(y_k,z_k)\}$ such that every column is beaten by some
element of $R$. Let $Y=(y_1,\dots,y_k)$.

\begin{claim}
For every $x\in L\cap\bit^n$ and every $z'$, some $i\le k$ satisfies
$V(x,y_i,z')=1$.
\end{claim}
\begin{proof}[Proof of the claim]
Consider the column $(x,z',y_n)$. Since $x\in L$, $V(x,y_n,z_i)=1$ for all $i$, so
the second alternative fails for every $i$. Hence the first holds for some $i$.
\end{proof}

\emph{Stage 2: a collective no-witness.} Define
\[
 W(x,z)=1\quad\Longleftrightarrow\quad V(x,y_i,z)=0\ \text{ for all }i\le k,
\]
a predicate computable in polynomial time given $Y$. By the claim, $W(x,z)=0$ for
all $z$ if $x\in L\cap\bit^n$. If $x\in\bit^n\setminus L$, then $W(x,z_n)=1$,
because $z_n$ beats every row on $x$. Hence $x\notin L$ iff $\exists z\ W(x,z)=1$, for
$x\in\bit^n$, and~\eqref{eq:dominating} holds with $w^*=z_n$. Lemma~\ref{lem:collective}, with
$\alpha=Y$, gives a list $z^*_1,\dots,z^*_m$ such that every $x\in\bit^n\setminus
L$ satisfies $W(x,z^*_j)=1$ for some $j$.

\emph{Decoding.} The advice for length $n$ is $(Y,z^*_1,\dots,z^*_m)$, padded to a fixed
polynomial length. The decoder
rejects $x$ iff some $j\le m$ satisfies $V(x,y_i,z^*_j)=0$ for all $i\le k$. If
$x\notin L$, such a $j$ exists. If $x\in L$, then $W(x,z)=0$ for every $z$, so no
such $j$ exists. Both stages run in polynomial time with the oracle, and the
correctness argument applies to every oracle consistent with $\SSE$.
\end{proof}

\begin{corollary}[the oblivious chain]\label{cor:chain}
\[
 \OGTD[\prMA]\subseteq\OLtP\subseteq\OtP\subseteq\OGTD[\prSBP]\subseteq\P^{\prSBP}\cap\Ppoly .
\]
Moreover, $\P^{\prOMA}\subseteq\OtP\subseteq\OGTD[\prSBP]$ and
$\OGTD[\prStP]\subseteq\OGTD[\prSBP]$.
\end{corollary}

\begin{proof}
Theorems~\ref{thm:OPprMA} and~\ref{thm:O2}, \ref{fact:F2}, \ref{fact:F4}, Theorem~\ref{thm:prS2}
and Proposition~\ref{prop:OP}(a).
\end{proof}
The class $\OM$ of Section~\ref{sec:ultimate} lies at the bottom of this chain:
$\OM\subseteq\OGTD[\sOprMA]\subseteq\OGTD[\prMA]$.

\begin{remark}
The containments of Corollary~\ref{cor:chain} mirror the uniform chain
$\P^{\prOMA}\subseteq\P^{\prMA}\subseteq\LtP\subseteq\StP\subseteq\P^{\prSBP}$
(Figure~\ref{fig:mirror}). There is one asymmetry. $\StP\subseteq\P^{\prStP}$ holds trivially, but we do not know whether
$\OtP\subseteq\OGTD[\prStP]$. Theorem~\ref{thm:O2} therefore does not follow from
Theorem~\ref{thm:B} by a reduction; it uses the Goldreich--Wigderson lemma with approximate counting
directly, twice. Using instead the $\FP^{\prAM}$ procedure of~\cite{CR11} for matrices with an
irrefutable row would give only the weaker $\OtP\subseteq\OGTD[\prAM]$.
\end{remark}

\section{Open problems}\label{sec:open}

\begin{enumerate}
\item \emph{Lower bounds below $\OMIN$.} Does $\OM$, or even $\OGTD[\prOMA]$, contain a
language outside $\SIZE[n^k]$ for every $k$? For $\OM$, this would give fixed-polynomial
lower bounds for both $\P^{\sOprMA}$ and $\ZPP^{\ONP}$. For
$\OGTD[\prOMA]$, it would give such lower bounds for $\P^{\prOMA}$.
\item \emph{The role of the tester.} In $\OM$, the $\co\ONP$ tester certifies the
generated advice; it is used only for the containment in $\ZPP^{\ONP}$
(Theorem~\ref{thm:three}). Can it be dropped, i.e., is
$\OGTD[\sOprMA]\subseteq\ZPP^{\ONP}$? Can the certified advice always be made
unique, i.e., is $\OUM=\OM$?
A positive answer to the first question would also make the tester-free class $\OGTD[\sOprMA]$ lie inside all three deepest collapse classes. 
\item \emph{The three deepest collapse classes.} No containment among $\OMIN$,
$\P^{\sOprMA}$ and $\ZPP^{\ONP}$ is known. Since $\ONP$ is contained in the last two
classes, $\P^{\sOprMA}\subseteq\OMIN$ or $\ZPP^{\ONP}\subseteq\OMIN$ would imply
$\ONP\subseteq\OMIN$, which is open~\cite{HV26,HVW26}. Conversely, $\OMIN\subseteq\P^{\sOprMA}$ or
$\OMIN\subseteq\ZPP^{\ONP}$ would give fixed-polynomial lower bounds for these classes.
If $\NP\subseteq\Ppoly$, then $\ONP\subseteq\OMIN$ (see
Corollary~\ref{cor:L2-collapse}). By Theorem~\ref{thm:ONP}, $\ONP\subseteq\OGTD[\prSBP]$
unconditionally.
\item \emph{Symmetric alternation versus linear orders.} Is $\OtP\subseteq\OLtP$, i.e.,
$\OtP=\OLtP$? Is $\StP\subseteq\LtP$, i.e., $\StP=\LtP$? Is
$\OtP\subseteq\OGTD[\prStP]$?
\item \emph{Promise oracles.} Chakaravarthy and Roy \cite{CR11} asked whether $\P^{\prStP}\subseteq\StP$.
For $\P^{\prStP}$, Theorem~\ref{thm:B} improves their upper bound $\P^{\prAM}$ to $\P^{\prSBP}$.
For $\StP$, our earlier paper~\cite{HV26} shows that it contains $\P^{\prMA}$, via its
subclass $\LtP$ defined by Korten and Pitassi~\cite{KP24}.
However, the original question remains open.
\item \emph{Zero error.} We proved that $(\StP \subseteq)\;\P^{\prStP}\subseteq \P^{\prSBP}$. At the same time, Cai's result \cite{Cai07} shows $\StP \subseteq \ZPP^{\NP}$ (in fact, $\P^{\prStP}\subseteq\ZPP^\NP$ is sketched in \cite[Theorem~4.16]{CR11} and also follows from \cite{FIKU08}). This raises a natural question: what is the relationship between $\ZPP^{\NP}$ and $\P^{\prSBP}$? For example, can we strengthen our theorem to $\ZPP^{\NP} \subseteq \P^{\prSBP}$?

\item \emph{Parallel queries.} Can the procedure of Theorem~\ref{thm:gw} be made
non-adaptive? This would give $\StP\subseteq\P^{\prSBP}_{\|}=\BPPpath$~\cite{OS18,HV26}.
For minimum finding, i.e., for $\LtP\subseteq\P^{\prSBP}_{\|}$, this question was asked
in~\cite[Section~6]{HV26}. Either containment would imply $\P^{\NP}\subseteq\PP$~\cite{HHT97},
which is open and fails relative to an oracle~\cite{Bei94}. Hence a positive answer
requires non-relativizing techniques.
\end{enumerate}

\section*{AI methodology}
This paper started as a follow-up to our earlier paper~\cite{HV26}, where the question of the smallest
Karp--Lipton collapse was investigated and the earlier containment
$\LtP\subseteq\P^{\prSBP}$ was proved. We used GPT-6 Astra and Claude Opus~5.5 for assistance
at most stages of the preparation of this paper, so these tools materially affected all
sections. Namely, we shared our private notes with these tools, discussed proofs with
them in an unstructured manner, asked Astra-6 to verify the results, 
and asked Opus~5.5 to draft both the conference submission
and the preprint version (and also to keep them in sync) in a dialogue with us.
Opus~5.5 formalized, at our request, the definition of the
class $\sOprMA$ (even though it was somewhat straightforward
from the definitions of our previous classes) and
the precise definition of the $\OGTD$ classes
(after we had formulated it in principle).
We the authors take full
responsibility for the paper, including the correctness and originality of all content, which
we verified. The final text was edited by the human authors.

\section*{Acknowledgments}
We are indebted to Ryan Williams, with whom we recently finished our previous paper \cite{HVW26} on the
topic and discussed multi-stage computations while working on it. 
The first author would like to thank the Isaac Newton Institute for Mathematical Sciences, Cambridge, 
for support and hospitality 
during the programme `Logical Foundations of Complexity Theory',
where part of the work on this paper was undertaken.
This event was supported by EPSRC grant EP/Z000580/1.
We greatly appreciate an open discussion with Oliver Korten during the workshop LFCW02
in the framework of this programme.
We thank the scientific organizers (Sam Buss and Jan Kraj\'{\i}\v{c}ek) for doing their excellent job, and the other participants for contributing to the stimulating atmosphere of the event.
We also thank Scott Aaronson, Yaroslav Alekseev, and John Hitchcock 
for conversations on related topics.

\phantomsection

\appendix

\section{Other routes to the collapses}\label{app:routes}

The main text derives the collapse to $\OMIN$ (Corollary~\ref{cor:main-KL}) from the
collapse to $\P^{\prMA}$ of~\cite{CR11}, and the collapse to $\OM$
(Theorem~\ref{thm:OMcollapse}) from the collapses to $\OMIN$ and to $\P^{\sOprMA}$. This
appendix gives other routes: through approximate counting instead of the collapse
of~\cite{CR11} (Section~\ref{app:counting}), through the collapse to
$\ZPP^{\NP}$~\cite{BCG+96,KW98}, which avoids promise problems
(Section~\ref{app:promise-free}), and to $\OM$ without $\OMIN$
(Section~\ref{app:learning}). Section~\ref{app:relativization} discusses
relativization.

\subsection{Through approximate counting}\label{app:counting}

\begin{remark}\label{rem:counting}
The containment $\PH\subseteq\P^{\prMA}$ under $\NP\subseteq\Ppoly$, used in the proof
of Corollary~\ref{cor:main-KL}, can also be obtained from Theorem~\ref{thm:B} instead
of~\cite{CR11}. Under the hypothesis, $\PH=\StP$ by Sengupta's argument~\cite{Cai07},
and $\StP\subseteq\P^{\SSE}$ by Theorem~\ref{thm:B}. Moreover, $\SSE$ has an
Arthur--Merlin protocol~\cite{GS86}, which can be made perfectly
complete~\cite{FGM+89}. In it, Merlin's messages can be replaced by circuits for the
search version of $\SAT$, as in~\cite{AKSS95} and~\cite[Proposition~4.1]{HV26}. As in
the proof of Lemma~\ref{lem:soprmaKL}, the correct circuits then dominate every other
message on every input and every random string. This gives $\SSE\in\sOprMA$ under the
hypothesis, and even $\PH\subseteq\P^{\SSE}\subseteq\P^{\sOprMA}$. Thus the collapse
passes through Theorem~\ref{thm:B}, with a single fixed oracle problem. In particular,
Corollary~\ref{cor:main-KL}, the collapse $\PH=\P^{\sOprMA}$ of
Lemma~\ref{lem:soprmaKL}, and hence Theorem~\ref{thm:OMcollapse} can be proved
without the collapse of~\cite{CR11}.
\end{remark}

\subsection{Without promise problems}\label{app:promise-free}

Both routes to the collapse $\PH=\P^{\sOprMA}$ (through~\cite{CR11} and
Lemma~\ref{lem:soprmaKL}, or through Remark~\ref{rem:counting}) end with a promise
oracle. By Lemma~\ref{lem:sample}, $\P^{\sOprMA}\subseteq\BPP^{\ONP}$, so such an oracle
can be traded for randomness and a language oracle. The collapse of K\"obler and
Watanabe~\cite{KW98} reaches a class of this kind directly, without promise problems:
if $\NP\subseteq\Ppoly$, then $\PH=\ZPP^{\NP}\subseteq\BPP^{\NP}$. It remains to
derandomize $\BPP^{\NP}$ inside $\LtP$, and~\ref{fact:F3} makes this possible.
Lemma~\ref{lem:bppnp} is the case $A=\SAT$ of~\cite[Lemma~6.12]{HVW26}, which states
$\BPP^{A}\subseteq\P^{A,\OLtP}$ for $A\in\Ppoly$, combined with
Lemma~\ref{lem:closure}; we include the short proof to keep the paper self-contained
for Definition~\ref{def:L2}.

\begin{lemma}[cf.~{\cite[Lemma~6.12]{HVW26}}]\label{lem:bppnp}
If $\NP\subseteq\Ppoly$, then $\BPP^{\NP}\subseteq\LtP$.
\end{lemma}

\begin{proof}
Let $L\in\BPP^{\SAT}$ via an oracle machine $M$ running in time $t(n)$, so that
$\Pr_r[M^{\SAT}(x,r)=L(x)]\ge3/4$ for every $x$. By hypothesis, $\SAT\cap\bit^m$
has circuits of size at most $m^d+d$ for a constant $d$. Fix $x$ of length $n$.
Replacing the oracle gates of the natural $\SAT$-oracle circuit for
$r\mapsto M^{\SAT}(x,r)$ by circuits for $\SAT$ yields an ordinary circuit with $t(n)$
inputs and size at most $s(n)$, for a polynomial $s$ independent of $x$. Let
$N=\max(s(n),t(n),8)$, and let $g_1,\dots,g_m$ be the list of \ref{fact:F3} on $1^N$,
computed with oracle $B\in\OLtP$, with $t(n)$-bit prefixes $g'_i$. Then
$\Pr_i[M^{\SAT}(x,g'_i)=L(x)]\ge3/4-1/8>1/2$. So $L(x)$ is the majority of
$M^{\SAT}(x,g'_i)$, which is computable in $\P^{\SAT,B}\subseteq\LtP$ by \ref{fact:F1}, \ref{fact:F2} and
Lemma~\ref{lem:closure}. The circuit for $\SAT$ is used only in the analysis.
\end{proof}

\begin{corollary}[cf.~{\cite[Corollary~2]{HV26}}]\label{cor:L2-collapse}
If $\NP\subseteq\Ppoly$, then $\PH=\LtP$ for Definition~\ref{def:L2}, and
$\ONP\subseteq\OLtP$.
\end{corollary}

\begin{proof}
By~\cite{KW98} and Lemma~\ref{lem:bppnp},
$\PH=\ZPP^{\NP}\subseteq\BPP^{\NP}\subseteq\LtP\subseteq\PH$. Hence, for every
$L\in\PH$, the tally language $T_L$ of Lemma~\ref{lem:canonical} is in $\LtP$, and
$L\in\P^{\TALLY\cap\LtP}=\OLtP$ by Theorem~\ref{thm:OPL2}. Thus $\PH=\OLtP$; in
particular, $\ONP\subseteq\NP\subseteq\OLtP$.
\end{proof}

\begin{remark}\label{rem:promise-free}
\begin{enumerate}
\item Corollary~\ref{cor:L2-collapse} also follows from Corollary~\ref{cor:main-KL},
but its proof is a second proof of Theorem~\ref{thm:ominKL}. It uses neither the
collapse of~\cite{CR11} nor Theorem~\ref{thm:OPprMA}: all its oracles are languages
($\SAT$ and the language $B$ of~\ref{fact:F3}), so the tally argument of
Theorem~\ref{thm:OPL2} applies directly. The proof of Corollary~\ref{cor:main-KL}
gives, in addition, $\PH=\OGTD[\prMA]$.
\item Unconditionally, it is open whether $\ONP\subseteq\OLtP$~\cite{HV26,HVW26}. By
Corollary~\ref{cor:L2-collapse}, a negative answer would imply
$\NP\not\subseteq\Ppoly$. For $\NP$ in place of $\ONP$, the question is equivalent to
$\NP\subseteq\Ppoly$ (Corollary~\ref{cor:equiv}).
\end{enumerate}
\end{remark}

\subsection{The collapse to \texorpdfstring{$\OM$}{OM} without \texorpdfstring{$\OMIN$}{OMIN}}\label{app:learning}

The collapse to $\OM$ (Theorem~\ref{thm:OMcollapse}) can also be proved without
$\OMIN$, along the lines of the proof of the collapse to $\ZPP^{\NP}$~\cite{BCG+96,KW98}.
The advice is a stack of $\SAT$-search
circuits, one for each quantifier block, in which each circuit may use the previous
ones. A failure of a circuit in the stack has a short witness, so the advice is
certified in $\co\NP$, which equals $\co\ONP$ under the hypothesis. The generator
finds the lexicographically first certified stack by a prefix search with a
$\SigmakP2$ oracle, which is in $\P^{\sOprMA}$ under the hypothesis. This proof
gives a unique generator output but not a unique accepted advice.

\subsection{Relativization}\label{app:relativization}

Lemmas~\ref{lem:closure}, \ref{lem:tally}, \ref{lem:ppoly}, \ref{lem:canonical},
Proposition~\ref{prop:OP} and Theorem~\ref{thm:transfer} relativize. Hence
Corollary~\ref{cor:main-KL} relativizes whenever the cited ingredients \ref{fact:F3}
and the collapse of~\cite{CR11} do; their proofs consist of circuit constructions that work for oracle circuits as
well (one should, however, carefully recheck all the dependencies for being relativizable).
For Theorem~\ref{thm:ominKL}, \ref{fact:F3} suffices: its second proof
(Corollary~\ref{cor:L2-collapse}) replaces the collapse of~\cite{CR11} by that
of~\cite{KW98}, whose proof relativizes, and Lemma~\ref{lem:bppnp} relativizes
whenever \ref{fact:F3} does. We do not use these facts.

\section{Further consequences}\label{sec:conseq}

\subsection{Derandomizing approximate counting}

We say that \emph{approximate counting is in $\FP^{\NP}$} if $\SSE$ is consistent
with a language in $\P^{\NP}$. Equivalently, given a circuit $C$ and a rational
$\eps>0$, a $(1+\eps)$-approximation of $\#C$ is computable with an $\NP$ oracle in time
polynomial in $|C|$ and $1/\eps$ (Lemma~\ref{lem:count}).

\begin{corollary}\label{cor:derand}
Suppose that approximate counting is in $\FP^{\NP}$. Then:
\begin{enumerate}
\item[\textup{(i)}] $\P^{\prStP}=\P^{\NP}$; in particular, $\StP=\P^{\NP}$;
\item[\textup{(ii)}] $\OtP\subseteq\OGTD[\NP]=\P^{\TALLY\cap\P^{\NP}}$, i.e., every
language in $\OtP$ has polynomial-size circuits constructible in $\FP^{\NP}$;
\item[\textup{(iii)}] for every $k$, some language with polynomial-size circuits
constructible in $\FP^{\NP}$ has no circuits of size $n^k$; in particular,
$\P^{\NP}\not\subseteq\SIZE[n^k]$;
\item[\textup{(iv)}] if moreover $\NP\subseteq\Ppoly$, then $\PH=\OGTD[\NP]=\P^{\NP}$.
\end{enumerate}
\end{corollary}

\begin{proof}
Let $A\in\P^{\NP}$ be consistent with $\SSE$. Proposition~\ref{prop:OP}(d), with $M$
the identity machine, gives $\P^{\SSE}\subseteq\P^{A}\subseteq\P^{\NP}$ and
$\OGTD[\SSE]\subseteq\OGTD[A]$. By Proposition~\ref{prop:OP}(b),
$\OGTD[A]=\P^{\TALLY\cap\P^A}\subseteq\P^{\TALLY\cap\P^{\NP}}=\OGTD[\NP]$. Now (i) follows
from Theorem~\ref{thm:prS2} and $\P^{\NP}\subseteq\LtP\subseteq\StP$, and (ii)
from Theorem~\ref{thm:O2}. For (iii), $\OLtP\subseteq\OtP\subseteq\OGTD[\NP]$ by (ii),
and $\OLtP\not\subseteq\SIZE[n^k]$ by \ref{fact:F2}. For (iv), $\PH=\OLtP\subseteq\OGTD[\NP]\subseteq\P^{\NP}$ by
Corollary~\ref{cor:main-KL}.
\end{proof}

\begin{remark}
Shaltiel and Umans~\cite{SU06} observed that Cai's algorithm~\cite{Cai07} is a
reduction to sampling. Hence it can be derandomized under a hardness assumption for
$\E^{\NP}$ that is implied by the one used to derandomize $\AM$, which gives
$\StP=\P^{\NP}$ under this assumption~\cite[Theorem~4.9]{SU06}. Item~(i) of
Corollary~\ref{cor:derand} requires only the derandomization of approximate
counting, and it also covers $\P^{\prStP}$. Conversely, Aydinlio\u{g}lu,
Gutfreund, Hitchcock and Kawachi~\cite{AGHK11} showed that derandomizing
approximate counting in this way implies exponential-size circuit lower bounds for
$\E^{\NP}$. Item~(iii) can be viewed as a fixed-polynomial, ``uniform-circuit'' analogue
of that implication.
\end{remark}

\subsection{Function classes, exponential time, and other classes inside
\texorpdfstring{$\StP$}{S2P}}\label{sec:func}

By Theorem~\ref{thm:prS2}, $\FP^{\prStP}\subseteq\FP^{\prSBP}$. In particular, every
search problem with a single-valued $\FP^{\prStP}$ algorithm, for instance a search problem whose
output-bit language\linebreak$\{(x,i):\text{bit $i$ of the output on $x$ is }1\}$ is in $\StP$, has a single-valued
$\FP^{\prSBP}$ algorithm. By a standard padding argument, Theorem~\ref{thm:B} scales
up to exponential time: $\StE\subseteq\E^{\prSBP}$, where in $\E^{\prSBP}$ the
oracle queries may have length $2^{O(n)}$. Together with the near-maximum lower
bound for $\StE$~\cite{CHR24,Li24} this gives
$\E^{\prSBP}\not\subseteq\SIZE[2^n/n]$, which was proved in~\cite{HV26} via
$\LtE\subseteq\E^{\prSBP}$. Finally, Theorem~\ref{thm:B} applies to every class
known to lie in $\StP$. For instance, the classes $\class{PTW}$ (Polynomial Tournament Winner),
$\class{PCW}\subseteq\class{PTW}$ (Polynomial Condorcet Winner) and $\class{PMA}$
(Polynomial Majority Argument) of unambiguous $\SigmakP2$ problems are contained in
$\StP$~\cite[Theorems~3.3, 4.2 and~5.3]{GGKN25}, and hence in $\P^{\prSBP}$. Since
$\LtP\subseteq\class{PTW}$~\cite[Theorem~3.4]{GGKN25} (their $\LtP$ is defined by linear
orders, as ours), already the containment $\class{PTW}\subseteq\P^{\prSBP}$ strengthens
$\LtP\subseteq\P^{\prSBP}$~\cite{HV26}. Similarly, approximating the value of a succinctly
given zero-sum game within an additive error is complete for
promise-$\StP$~\cite{FIKU08}. Hence such approximations,
obtained by binary search, can be computed in $\FP^{\prSBP}$.

\subsection{Minimum finding}\label{app:min}

As a special case, the Goldreich--Wigderson lemma with approximate counting
(Theorem~\ref{thm:gw}) gives another minimum-finding algorithm, and thus another
proof of $\LOP\in\FP^{\prSBP}$ and $\LtP\subseteq\P^{\prSBP}$~\cite[Theorem~3.14]{HV26}.

\begin{corollary}\label{cor:min}
Given a circuit $E$ that defines a strict linear order $\prec$ on $\bit^a$, the
minimum of $\prec$ is computable in $\FP^{\SSE}$; the output is the same for every
oracle consistent with $\SSE$. Consequently, $\LOP\in\FP^{\prSBP}$.
\end{corollary}

\begin{proof}
Consider the matrix $V(u,w)=1$ iff $u=w$ or $u\prec w$. Its minimum $\mu$ is an
irrefutable row, so Corollary~\ref{cor:irrefutable} gives $R$ such that every column
is beaten by some row of $R$. For the column $\mu$, some $u\in R$ satisfies $u=\mu$
or $u\prec\mu$; hence $\mu\in R$, and $\mu$ is the $\prec$-least element of $R$,
which we output. For $\LOP$, given an arbitrary circuit $E$, first search for a
violation of the axioms of strict linear orders (irreflexivity, totality,
asymmetry, transitivity) using $\SAT\in\P^{\SSE}$ and search-to-decision; if there is
none, output the minimum.
\end{proof}

\end{document}